\documentclass[aps,twocolumn,amsmath,letterpaper]{revtex4}
\usepackage{amssymb}
\usepackage{graphicx}
\usepackage{array}
\usepackage{hhline}
\usepackage{longtable}
\cleardoublepage

\renewcommand{\thetable}{\Roman{table}} \thetable

\newcommand{\vetor}[1]{\mbox{\boldmath ${#1}$}}

\begin{document}

\title{ Finite-Connectivity Spin-Glass Phase Diagrams and Low Density
Parity Check Codes }

\author{Gabriele Migliorini and David Saad}
\affiliation{ The Neural Computing Research Group\\ School of
Engineering and Applied Sciences, Aston University\\ Birmingham B4
7ET, United Kingdom.}

\begin{abstract}
We obtain phase diagrams of regular and irregular finite
connectivity spin-glasses. Contact is firstly established between 
properties of the phase diagram and the performances of low density
parity check codes (LDPC) within the Replica Symmetric (RS) ans\"atz.
We then study the location of the dynamical and critical transition of these
systems within the one step Replica Symmetry Breaking theory (RSB), extending
similar calculations that have been performed in the past for the
Bethe spin-glass problem. We observe that, away from the Nishimori
line, in the low temperature region, the location of the dynamical
transition line $does$ change within the RSB theory, in comparison 
with the (RS) case. For LDPC decoding over
the binary erasure channel we find, at zero temperature and rate $R=1/4$ an RS
critical transition point located at $p_c \simeq 0.67$ while the
critical RSB transition point is located at $p_c \simeq 0.7450 \pm
0.0050$, to be compared with the corresponding Shannon bound $1-R$. For the
binary symmetric channel (BSC) we show that the low temperature reentrant
behavior of the dynamical transition line, observed
within the RS ans\"atz, changes within the RSB theory; the location of
the dynamical transition point occurring at higher values
of the channel noise. Possible practical implications to improve the
performances of the state-of-the-art error correcting codes are discussed.\\


\end{abstract}
\maketitle
\def\s{\rule{0in}{0.28in}}
\section{Introduction}

\setlength{\LTcapwidth}{\columnwidth} The survival of
ferromagnetic ordering under the disruption of frozen random
fields~\cite{IM} and the onset of Spin-Glass (SG) order in systems
characterized by random competing interactions~\cite{EA} are two
central problems in the statistical mechanics of systems with
quenched randomness. When properly rephrased, these two problems
turn out to relate to core problems in other, quite different
disciplines. Insight into these two closely related problems in
any given system is obtained from the corresponding phase diagram
as a function of the physical parameters, such as temperature and
disorder strength.

Finite-connectivity mean-field SG models~\cite{WS,MP1,Mo,DM,DG}
and their {\em p}-spin counterpart~\cite{RK} are important for two
distinct reasons. Firstly, despite being mean-field in nature,
they are believed to share common properties with
finite-dimensional spin-glass systems~\cite{Mon,GL}. Secondly, the
statistical mechanics of SG model systems with fixed finite
connectivity relates to the study of various hard computational
problems. In the last two decades the analysis of
finite-connectivity SG systems has also offered new tools for
understanding the very nature of several optimization problems
{\em and} improve algorithmic performances~\cite{MP2,MPZ}. In this
paper we first obtain the phase diagram of finite-connectivity
mean-field SG models. We then investigate the relation between
features of the phase diagram and the performance of Low Density
Parity Check (LDPC) Codes, analyzing the effects induced by the
Replica Symmetry Breaking (RSB) nature of solutions obtained for
these finite-connectivity SG models~\cite{Mo2}. In order to answer
these questions, we exploit the relation between
finite-connectivity SG systems and a class of exactly solvable
models; we employ an efficient computational method that has been
shown to be quite effective for understanding several features of
finite-dimensional SG~\cite{MB}.

We will consider a generalized model unifying the frameworks 
for a few distinct problems.

Particular attention will be given to two special cases: a) The Bethe SG
which has been extensively studied under both Replica Symmetric
(RS) and RSB  ans\"{a}tze \cite{WS,MP1,Mo} and recently 
considered in its ferromagnetically biased
version \cite{Castellani}. b) Gallager (LDPC) error-correcting codes~\cite{Gallager}. 
While the former is well known within the statistical physics community, many in this
community are less familiar with the latter and its links to the
physics of disordered systems.

Reliable transmission of information in noisy conditions is one of
the basic problems of modern communication technology.
Correspondingly, one of the central problems of information theory
is channel coding. Methods to achieve reliable information
transmission rely on the introduction of structured redundancy to
the original message in order to compensate for corruption due to
noise. Shannon derived rigorous bounds on the level of redundancy 
(the code rate) required to enable error-free communication for a given 
channel noise~\cite{Shannon}.
However, Shannon's theorems are non-constructive and do not offer
a practical coding scheme. Different methods, based on various
redundancy construction recipes, have been proposed over the
years~\cite{book}. Despite the fact that error-correcting codes
are now widely used in a variety of applications, current
performances of most methods are significantly below Shannon's
bound. One family of error-correcting codes that have been shown
to provide close-to-optimal performances is that of Gallager's
LDPC codes~\cite{Gallager}. In these codes the structured
redundancy is introduced through parity checks of Boolean sums of
randomly sampled message bits. The relation between the
parity-check error-correcting codes and SG models, and
consequently between quite different disciplines as statistical
physics and information theory, has been pointed to in the seminal
work of Sourlas~\cite{Sourlas1}.

Decoding in LDPC codes corresponds to a class of SG models defined
by an underlying lattice geometry that reflects the low density
character of the code construction. The importance of this very
last point, that made the above correspondence useful and of
practical relevance was not understood until recently~\cite{Saad}.
Methods developed in the statistical mechanics of diluted SG under
the RS ans\"atz have been successfully employed to compute
macroscopic properties of such systems for different parameter
values; reasonable agreement with observed decoding bounds have
been reported~\cite{Vicente}. To our knowledge however, the
only phase diagram that has been evaluated for LDPC codes is in
the limit of large coordination/large multi-spin interaction
numbers~\cite{Montanari}, which also relates to the phase
diagram of the Random Energy Model (REM)~\cite{Derrida}.

In the following section we introduce the class of Ising spin systems
to be analyzed and its links to the Bethe SG and LDPC decoding
problems.  This will be followed by section~\ref{sec:RS} where we 
review the RS equations and the corresponding solutions are obtained 
using a computational method adopted from the study of 
$d$-dimensional hierarchical models. As part of this
section we will present the method and briefly discuss the relation
between finite-connectivity SG and $d$-dimensional hierarchical
models.  
In section~\ref{sec:RSB} we introduce the equations
behind the RSB theory of finite-connectivity SG studied here and present results
obtained for the Bethe SG problem  and LDPC 
error-correcting codes under two different transmission channels: 
the binary erasure channel (BEC) and the binary symmetric channel (BSC).
In section~\ref{sec:disc} we discuss the results 
obtained and point to future research directions.

\section{The Model}
\label{sec:model}

The two different models examined in this work stem from a unified
physical system of $K$-spin interaction Ising model on a lattice of
fixed connectivity $C$. The Hamiltonian describing the system takes the
form
\begin{equation}
 { \cal H}_{0}= - J \sum_{\mu} {\cal A}_{ \mu}
 \prod_{i \in { \cal L}(\mu) } S_i \ ,
 \label{H_0}
\end{equation}
where $\mbox{ \boldmath $S$} \in \{ \pm 1 \}^{N} $ represents $N$
binary variables, $ {\cal L}(\mu)=\langle i_1, \ldots ,i_K \rangle$
a set of indices related to interaction $\mu$ and $ {\cal A}_{
\mu}$ the corresponding connectivity tensor with the following
properties:
\begin{eqnarray}
{\cal A}^{ \mu}_{ \langle i_1, \ldots ,i_K \rangle } &=& \left\{
 \begin{array}{cc} 1 & \mbox{for}~~~ {\cal L}( \mu)=\langle i_1,
 \ldots i_K \rangle \\ 0 & \mbox{otherwise} \end{array}
 \right. \nonumber \\
\sum_{\langle i_2, \ldots, i_K \rangle } {\cal A}_{ \langle i,i_2,
\ldots, i_K \rangle } &=& C ~~~ \forall i \ .
\label{tensor}
\end{eqnarray}
The parameter $J$ is related to the strength of the random exchange
interaction. 

The various models considered in this work differ in the
connectivity parameters and type of bias used, which corresponds to an
additional component of the Hamiltonian~(\ref{H_0}).

\subsection{The Bethe SG}

The Bethe SG has been extensively studied in the past~\cite{WS,MP1,Mo}
and represents a special case of the Hamiltonian (\ref{H_0}) where one
is only considering pairwise interactions, $K=2$.

Bethe SG with ferromagnetically biased exchange interaction $J$, which
we will focus on here, has been recently studied in~\cite{Castellani},
where the distribution of the quenched random variable $J$ takes the
form
\begin{equation}
P(J)= \frac{1 + \rho}{2}~ \delta(J-1)+ \frac{1 - \rho}{2}~ \delta(J+1) \ ,
\end{equation}
and where the parameter $ \rho \in [0,1]$ is $ \rho=1$ for a
ferromagnet and $ \rho=0$ for the unbiased Bethe SG.

\subsection{LDPC error-correcting codes}
\label{sec:LDPC}

We consider a simple communication model whereby $L$ bit messages are
encoded using LDPC codes and then communicated through two channel
types: $a)$ The Binary Symmetric Channel (BSC) where bits are flipped
with probability $p$ during transmission and $b)$ The Binary
Erasure Channel (BEC) where bits are received with probability $1-p$
and fail to arrive with probability $p$.

A Gallager LDPC code is defined by a binary matrix
$\mbox{\boldmath $H$}=[\mbox{\boldmath $A$}\mid\mbox{\boldmath
$B$}]$, concatenating two very sparse matrices known to both
sender and receiver, with $\mbox{\boldmath $B$}$ (of
dimensionality $(N-L)\times(N-L)$) being invertible; the matrix
$\mbox{\boldmath $A$}$ is of dimensionality $(N-L)\times L$. LDPC
codes can be employed over various finite fields \cite{nakamura}; 
we will restrict the treatment in this paper to binary symbols.

Encoding refers to mapping the original $L$ dimensional binary
message vector $\mbox{\boldmath $\xi$}\in\{0,1\}^L$ to an $N$
dimensional binary code word $\mbox{\boldmath $t$}\in\{0,1\}^N$
($N>L$) by taking the product $\mbox{\boldmath
$t$}=\mbox{\boldmath $G^T$}\mbox{\boldmath $\xi$}\;\mbox{(mod
2)}$, where all operations are performed in the field $\{0,1\}$
and are indicated by $\mbox{(mod 2)}$. The generator matrix is
$\mbox{\boldmath $G^T$}= [\mbox{\boldmath $I$}\mid ( \mbox{\boldmath
$B^{-1}$}\mbox{\boldmath $A$})^{T}] \mbox{ (mod 2)}$, where $\vetor{I}$
is the $L\times L$ identity matrix, implying that $\mbox{\boldmath
$H$}\mbox{\boldmath $G^T$} \mbox{ (mod 2)} =0$ and that the first
$L$ bits of $\mbox{\boldmath $t$}$ are set to the message
$\mbox{\boldmath $\xi$}$. In {\it regular} Gallager codes the
number of non-zero elements in each row of $\vetor{H}$ is chosen
to be exactly $K$.  The number of elements per column is then
$C=(1-R)K$, where the code rate is $R=L/N$ (for unbiased
messages). We will mostly focus here on regular code
constructions.

\subsubsection{BSC}

The corruption process that is most intuitively
linked to physical systems is the BSC, where the encoded vector
$\mbox{\boldmath $t$}$ is corrupted by flip noise represented by
the vector $\mbox{\boldmath $\zeta$}\in\{0,1\}^L$ with components
independently drawn from
$P(\zeta)=(1-p)~\delta(\zeta)+p~\delta(\zeta-1)$.  The received
vector takes the form $\mbox{\boldmath $r$}=\mbox{\boldmath
$G^T$}\mbox{\boldmath $\xi$}+\mbox{\boldmath $\zeta$}\mbox{ (mod
2)}$.

Decoding is carried out by multiplying the received message by the
matrix $\vetor{H}$ to produce the {\it syndrome} vector
$\mbox{\boldmath $z$}=\mbox{\boldmath $H$}\mbox{\boldmath
$r$}=\mbox{\boldmath $H$}\mbox{\boldmath $\zeta$}\mbox{ (mod 2)}$
from which an estimate $\mbox{\boldmath $\widehat\zeta$}$ for the
noise vector can be inferred. An estimate for the original message
is then obtained as the first $L$ bits of $\mbox{\boldmath $r$}
+\mbox{\boldmath $\widehat\zeta$}\mbox{ (mod 2)}$.

The Bayes optimal estimator (also known as {\it Marginal Posterior
Maximiser} (MPM)) for noise vector bits is defined as
$\widehat{\zeta}_j=\mbox{argmax}_{S_j}P(S_j\mid \mbox{\boldmath
$z$})$.  Computing the MPM estimate directly is computationally
hard, and is typically approximated in practice by employing an
iterative inference algorithm such as belief propagation~\cite{MacKay}.

The connection to statistical physics becomes clear when the field
$\{0,1\}$ is replaced by Ising spins $\{\pm 1\}$ and mod $2$ sums
by products~\cite{Sourlas1}. The syndrome vector acquires the form
of a multi-spin coupling ${\cal J}_\mu=\prod_{i\in{\cal
L}(\mu)}\zeta_i$ where $i=1, \ldots, N$ and $\mu=1, \ldots,
(N-L)$. The $K$ indices of nonzero elements in the row $\mu$ of
$\vetor{A}$ are given by ${\cal L}(\mu)=\{ i_1, \ldots, i_K \}$ 
and where ${ \cal A}_{ \langle i_1, \cdots i_K \rangle }$ is the 
connectivity tensor is given in equation (\ref{tensor}).
After gauging $S_j \rightarrow S_j \zeta_j$ the multi-spin
coupling coefficient vanishes $({\cal J}_\mu=1)$ and the parity
check component of the Hamiltonian takes the form (\ref{H_0}) with
$J\rightarrow \infty$ forcing all candidate vectors to obey the
parity checks. This notation is used to
distinguish the temperature from the parity check constraints and to 
facilitate the finite temperature study of error-correcting codes.

Prior knowledge of the biased noise vector is added through a
second component to the Hamiltonian (\ref{H_0}). After gauging,
the new term takes the form $H_1= - F \sum_{j=1}^N S_j \zeta_j$
where $F =\frac{1}{2} \log( \frac{1-p}{p})$ is the Nishimori
condition~\cite{Nishi1,Nishi2} (where $\beta=1$), which
corresponds to the correct prior for the variables
$\mbox{\boldmath${S}$}$. Decoding at the Nishimori condition
corresponds to the MPM estimation~\cite{Iba}. This gives rise to
the Hamiltonian
\begin{equation}
-\beta { \cal H}= \beta J \sum_{\mu} {\cal A}_{ \mu}
 \prod_{i \in { \cal L}(\mu) } S_i + \beta F \sum_{j=1}^N S_j \zeta_j\ ,
 \label{RFRB}
\end{equation}
for the BSC.

Another important estimator one should mention is the {\it Maximum A Posteriori}
(MAP) estimator defined as $\mbox{\boldmath$
\widehat{\zeta}$}=\mbox{argmax}_{\mbox{\boldmath $S$}}
P(\mbox{\boldmath $S$} \mid \mbox{\boldmath $z$})$. This
corresponds to zero temperature decoding~\cite{Iba}.

\subsubsection{BEC}

Codeword bits transmitted over a BEC have a probability $1-p$ of
being received (in tact). Encoding is carried out similarly to the
BSC, but decoding is based on inferring the codeword itself,
rather than the noise vector.

In the case of the BEC, solution vectors are, by definition,
codewords and should therefore obey $\mbox{\boldmath
$H$}\mbox{\boldmath $S$}\mbox{ (mod 2)} =0$, where
$\mbox{\boldmath $S$}$ are the received vectors (with a certain
number of unknown bits).

An estimate of the original message is then obtained by inferring
the missing bits using MPM. Similar 
practical methods to those described in the BSC may be employed
also in this case.

The statistical mechanics formulation of the problem is 
similar to that of the BSC, except for the second component added to the
Hamiltonian (\ref{H_0}), that takes the form
\begin{equation}
H_1 = - \sum_{i=1}^{N} \ln \left[p + (1-p) \prod_i \delta_{S_i,1}\right] \
.
\end{equation}
Code performance in the BEC will be considered only at the zero
temperature limit as in~\cite{Franz}.

A review of  LDPC codes and their link to statistical mechanics
has  been presented in~\cite{KSreview}.

\section{Replica Symmetric Theory} 
\label{sec:RS}

The analysis carried out here is a natural extension of similar
calculations done in the past for the Bethe SG~\cite{BL} model and
$p$-spin Ising models~\cite{WS}. We therefore briefly review the
RS theory of the model described by Hamiltonian (\ref{RFRB}) as
both the Bethe SG and LDPC decoding can be considered as special
cases of the same model.

To compute the free energy of the problem, consider the replica equality,
\begin{equation}
\beta f = - \lim_{N \rightarrow \infty} \frac{1}{N} \frac{ \partial }{
\partial n} \Big |_{n \rightarrow 0} \langle {\cal Z}^n \rangle_{ {
 \cal A}, \zeta,J},
\end{equation}
where $ \langle {\cal Z}^n \rangle_{ { \cal A}, \zeta,J}$ is the
analytical continuation in the interval $n \in [0,1]$ of the
replicated partition function~\cite{MPV}. Following the standard
derivation, e.g. as in~\cite{Vicente,MKSV}, and employing the RS
ans\"atz, one obtains the following expression for the replica
symmetric free-energy  as a function of  the local field
distribution $\pi(x)$ (details can be found in Appendix A)
\begin{equation}
 \beta f = \Delta f^{(1)} - C \frac{K-1}{K} \Delta f^{(2)}.
\end{equation}
There are both site and bond contributions,
\begin{eqnarray}
\Delta f^{(1)}  &=& -  E_{J, \zeta} \int \prod_{j=1}^C
\prod_{j'=1}^{K-1}
 dx_{jj'} ~ \pi(x_{jj'})~ { \cal F}^{(1)}( \{ x_{jj'}\},\zeta,J_j) \nonumber \\
\Delta f^{(2)}  &=& -  E_{J}   \int \prod_{j=1}^K dx_j ~ \pi(
 x_j)~ { \cal F}^{(2)}( \{ x_{j}\},J_j),
\label{free-energy}
\end{eqnarray}

where $E_{J, \zeta}$ denotes an average over the two sources of
randomness in (\ref{RFRB}), and
\begin{eqnarray}
{ \cal F}^{(1)}_a&=& \log 2 \cosh  \beta \Big ( \sum_j u( \{
  x_{jj'}\},J_j)+ F \zeta \Big ) \nonumber \\
{ \cal F}^{(1)}_b&=&a( \{x_{jj'}\},J_j ),\nonumber \\
{ \cal F}^{(2)} &=& a( \{ x_j \},J_j)+u( \{ x_j \},J_j) \ .
\end{eqnarray}
We denote ${ \cal F}^{(1)}={ \cal F}^{(1)}_a+{ \cal F}^{(1)}_b$
where
\begin{eqnarray}
\beta u( \{ x_j \},J)&=& \frac{1}{2} \log [ R(+)/R(-)] \nonumber \\
\beta a( \{ x_j \},J)&=& \frac{1}{2} \log [ R(+)R(-) ] \ .
\end{eqnarray}
The polynomials $R(\sigma)$ are defined via,
\begin{equation}
R(\sigma)= \sum_{ \sigma_1, \cdots \sigma_{K-1}}
\exp \beta [ y_1 \sigma \sigma_1 + y_2 \sigma_1 \sigma_2 +
\cdots y_K \sigma_{K-1} \sigma ],
\label{poly}
\end{equation}
where $\{ y_1, \cdots, y_K \} \equiv \{ x_1, \cdots x_{K-1},J \}$.
The above expression, in a similar form, has already been
obtained in ~\cite{Saad}, and the equivalence with their expression is
shown in Appendix A. We prefer the current formulation because of
its manifest simplicity (the two contributions to the free-energy
coming from sites and bonds are here in evidence) and because it
depends on the function $a( \{ x_j \},J_j)$, which will play a
central role in the RSB case, examined in section~\ref{sec:RSB}.

The above writing also introduces the polynomials $R(\sigma)$,
already well known in the context of hierarchical
lattices~\cite{MB}. Evaluating the saddle point equations of the
above free-energy expression is the basic premise of the RS
theory, so that the crux of the quenched randomness problem
(\ref{RFRB}) lies in the convolution~\cite{Saad},
\begin{equation}
\pi(x) = E_{J,\zeta} \int \prod_{ij} dx_{ij}~ \pi(x_{ij})~
 \delta \Big( x- { \cal R} ( \{x_{ij} \}, J, \zeta) \Big ),
\label{crux}
\end{equation}
where $i=1, \cdots,C-1$, $j=1, \cdots ,K-1$; and where the
recursion relations are defined via
\begin{equation}
{ \cal R} ( \{x_{ij} \}, J, \zeta) = \sum_{j=1}^{C-1} u (
 \{x_{ij}\},J_j)+ F \zeta \ .
\label{recursion}
\end{equation}
We can also rewrite for convenience the above convolution with the
aid of the auxiliary distribution $\hat{ \pi} ( \hat{x} )$, as a
set of two coupled equations (cfr. Appendix A),
\begin{eqnarray}
\pi(x)&=& E_{ \zeta} \int \prod_{i=1}^{C-1} d \hat{x}_i~ \hat{ \pi
}(\hat{x}_i
 )~ \delta \Big ( x - \sum_{i=1}^{C-1} \hat{x_i}- F \zeta \Big) \nonumber \\
\hat{ \pi }( \hat{x} ) &=& E_J \int \prod_{j=1}^{K-1} dx_j ~\pi(x_j)~ \delta
\Big( \hat{x}-u( \{ x_j\},J) \Big).
\label{crux2}
\end{eqnarray}

\subsection{Saddle Point Equations and LDPC Decoding}
To briefly explain the link between LDPC decoding and the replica
formulation we will restrict this discussion to the BSC. The
replica analysis as well as decoding in other noisy channels
follow a similar path~\cite{tanaka}.

Following the derivation of~\cite{Vicente} we write saddle point
equations for the LDPC decoding problem, considering a special
case of the Hamiltonian (\ref{RFRB}) and the saddle point
equations (\ref{crux2}), where $J \rightarrow \infty$ is
deterministic 
\begin{eqnarray}
\pi(x)&=& E_{ \zeta} \int \prod_{i=1}^{C-1} d \hat{x}_i~ \hat{ \pi
}(\hat{x}
 )~ \delta \Big ( x - \sum_{i=1}^{C-1} \hat{x_i}- F \zeta \Big)
 \nonumber \\
\hat{ \pi }( \hat{x} ) &=&  \int \prod_{j=1}^{K-1} dx_j ~\ pi(x_j)~ \delta
\Big( \hat{x}-u_{ \beta }( \{ x_j \})  \Big)~.
\label{crux3}
\end{eqnarray}
Most of the numerical results obtained so far are obtained along
the Nishimori line $ \beta =1$. Here we investigate the
phase diagram at $any$ prior temperature $1/ \beta F$, for
different choices of the parameters $C$ and $K$. A typical case we
will discuss in detail is the one corresponding to $C=3$, $K=4$.

We will check the consistency between the results obtained and
true decoding experiments, using Belief Propagation (BP)~\cite{KF}, at
and away from the Nishimori condition. We will not review 
BP in the paper as it is a well known message
passing algorithm~\cite{MacKay,Pearl} with strong links to the
Bethe approximation and its variants~\cite{YFW,OS}. The iterative
algorithm is a microscopic equivalent of equations~(\ref{crux3}).

\subsection{The Computational Method}

The set of (RS) equations (\ref{crux2}) have been considered and
solved in a variety of different ways already. This section
introduces to a different method.
We propose an alternative approach to existing methods (e.g.
population dynamics or Monte Carlo techniques) which solves
equations of the form (\ref{crux}) efficiently and accurately. The
main reason why it has not been used to study finite-connectivity
SG models so far is because of the different context in which it was firstly 
introduced \cite{AB}.
The method we propose allows to determine the phase diagram of
diluted systems of the type studied here and analyze the nature
and exact location of the phase boundaries. It borrows tools and
concepts from a different branch of research, namely the area of
hierarchical models \cite{MB,HB}, to obtain accurate phase diagrams.
The iterative method we now
describe has a corresponding computational effort that scales as
${ \sf N}^{K-1}$, { \sf N} being the number of bins introduced and $K$ the
multi-spin interaction parameter.

Equation (\ref{crux}) complicates, after a few iterations, even a
simple initial local field distribution. The distribution $\pi(x)$
and its conjugate distribution $\hat{ \pi }(\hat{x} )$ are
represented via histograms. Each histogram is characterized by two
quantities, $x_{ij}$ and $\pi(x)$, where the latter is the associated
probability. Equation (\ref{crux3}) is the convolution of $ (K-1)
C$ distributions, taking into account the distribution of the
random field $P(\zeta)$, contained in the average $E_{ \zeta} (..)
= \int P( \zeta) d \zeta (..)$. This is numerically constructed
from pairwise convolutions of distributions. For a $R=1/4$ rate
Gallager code , e.g., three pairwise convolutions are required.
A pairwise convolution (or binary operation) is achieved as follows:\\
(1) The histograms are placed on a grid properly drawn at each iteration step of
the algorithm. All histograms that fall within the same grid cell are
combined in such a way as to preserve the averages across the domain of
the local field (and its conjugated) distribution(s). The histograms
that falls outside the grid, representing a pre-determined small
fraction of the overall probability, are similarly combined into a single
histogram. In this procedure, a histogram that falls within a narrow
band of a grid boundary is proportionally shared between neighboring
bins. The gridding is done separately for $x>0$ and $x<0$. \\
(2) Two gridded distributions are convoluted according to equation
(\ref{crux2}) regenerating the original number of histograms.\\
The convolution of $C(K-1)$ distributions is achieved, from
pairwise convolutions, as follows: (1) The pairwise convolution
contained in the first equation of (\ref{crux2}) is cycled once
for, e.g., a code rate $R=1/4$. (2) The resulting distribution is
pairwise-convoluted with the noise channel distribution $P(
\zeta)$. (3) The resulting distribution is further self-convoluted
according to the second equation in (\ref{crux2}). This completes
the entire process, that is then iterated until convergence is
achieved. Multiple cycling operations are also an interesting
option being related to, e.g., the source compression problem. The
flows of the probability distribution $\pi(x)$ determine the phase
diagram of the system. Most of our results have been obtained
considering $125 \times 4$ independent bins, corresponding to the
flows of $10^3$ quantities (the corresponding number of elementary 
operations considered is the number of bins to the 
power of $K-1$). Each of the two
grids at $x>0$ and $x<0$ have e.g. $125$ bins on each side of the 
average value.\\
Similar set of equations to (\ref{crux}) have been discussed in a
variety of different contexts. A very important one is that of
real-space renormalization-group theory~\cite{AB} \cite{HB} \cite{MKB}. 

Eventhough the algebraic form of the convolution corresponding to these 
two problems is the same, the two problems are indeed very different. In the 
case of hierarchical models the distribution of the exchange interaction
is studied, while in the problem we are discussing in this work one 
considers the distribution of the local field. 
This reflects the difference in the two underlying geometries. 
The $d$-dimensional hierarchical lattice include loops
that are one of the essential characteristics that
makes hierarchical lattices better candidates, within exactly
solvable models, to study the problem of finite-dimensional SG~\cite{MB}.
The similarity between equations~(\ref{crux3}) and those 
related to hierarchical models is indicative of a deeper relation 
between the two ( both geometries are hierarchical in the sense of 
ref. \cite{KG}) and will be the subject of future work.

\subsection{The LDPC Phase Diagram - BSC}

\begin{figure*}
\centering
\includegraphics*[scale=0.6]{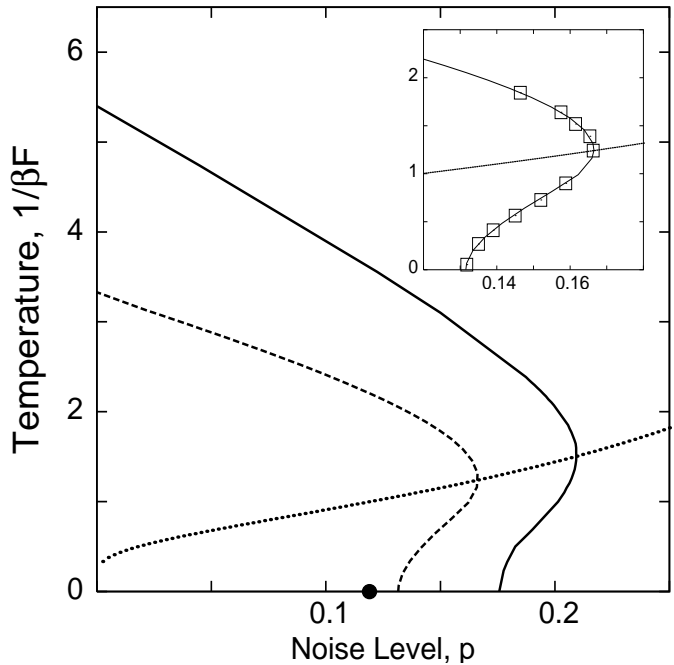}
\caption{The phase diagram for the LDPC code of rate $R=1/4$ over the
 BSC channel. The solid and dashed lines represent the critical and
 dynamical transitions, respectively. The inset shows the curvature of
 the dynamical transition line boundary, together with  decoding experiments
 results (square boxes). The dotted curve represents the Nishimori line. The black
 filled dot indicates the location of the dynamical transition at zero
 temperature assuming the local field distribution to be peaked over
 integer values only. 
\label{ldpc_phase}}
\end{figure*}

The calculated phase diagram for the Gallager LDPC code of rate $R=1/4$ 
and a BSC of flip rate $p$ as a function of the prior temperature $1/ \beta F$ 
are shown if Fig.~\ref{ldpc_phase}. We report both the location of
the dynamical noise transition $p_d$, defined by the noise level
at which suboptimal solutions emerge, and critical noise transition
$p_c$, where the ferromagnetic solution ceases to be dominant, for all ranges
of temperature $1/ \beta F$. The phase diagram was calculated
considering the flows of $10^3$ quantities dictated by the convolution (\ref{crux}).
As mentioned before the above number should be taken to the power $K-1$ to obtain
the number of elementary operations within, e.g., the second equation of (\ref{crux2}).

\begin{figure*}
\centering
\includegraphics*[scale=0.6]{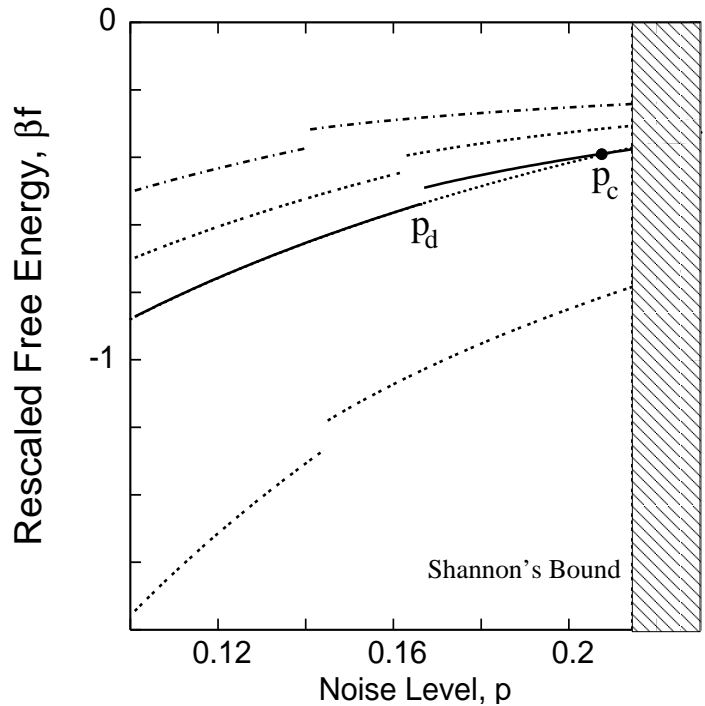}
\caption{ The computed free-energy (\ref{free-energy}) as a function of the noise
 level $p$ for different values of $ \beta$. The solid line corresponds
to the free-energy at the Nishimori line $ \beta=1$, while the
 dashed lines correspond to the values of $ \beta=8/6,8/4$ below the
 Nishimori line and $\beta=8/10,8/14$ above it. For flip rates $p$
 larger than the dynamical
 transition the computed free-energy is higher than the ferromagnetic
 one, given by $ f_{\mbox{Ferro}} = - \frac{1-2p}{2}log(\frac{1-p}{p}) $ (dotted-line). 
\label{bscRSfree}}
\end{figure*}

The location of the dynamical and critical transition values can
be determined by analyzing the free-energy and the complexity as a
function of the channel noise level for different values of the
temperature $1/ \beta$ (Fig.~\ref{bscRSfree}). We will refer noise
level that corresponds to the emergence of sub-optimal solutions
as the dynamical transition. At any value $p<p_d$, the computed
free-energy (\ref{free-energy}) reduces to the ferromagnetic
free-energy $f=- \frac{1-2p}{2} \log(\frac{1-p}{p})$. Consider first
temperature values in the proximity of the Nishimori line. The
trajectories exhibit a crossover to an unstable distribution 
first, and then flow to the ferromagnetic
sink below $\simeq 0.1665$ (corresponding to a decoding success) 
or to a stable non-ferromagnetic distribution (corresponding to a 
decoding failure) above it. 
A similar behavior is observed for any temperature above the Nishimori
line, the stable distribution being different for each values of
the temperature $ \beta< 1$. Rather different behavior is
observed below the Nishimori line, where the unstable distribution
is a strong coupling distribution, meaning that the average local
field strength increases but the rescaled distribution remains 
constant. Finally, the trajectory flows again, either to
the ferromagnetic sink or to a fixed local field distribution,
which is stable under iteration of the recursion (\ref{crux3}).

The computed free-energy, according to the convolution
(\ref{crux3}), is higher than the ferromagnetic free-energy for all
noise levels $p_d<p<p_c$. Above the critical noise value $p_c$,
the computed free-energy becomes lower than the ferromagnetic
free-energy. The location of the dynamical transition has
been determined, at any temperature, considering the flows, under
iteration, of the convolution (\ref{crux}), while the location of
the critical noise threshold has been determined from the
free-energy (\ref{free-energy}) as the point where the computed
and the ferromagnetic free-energies coincides. In the inset of
Fig.~\ref{ldpc_phase} we also report in square boxes the results 
of real decoding experiments obtained as follows. We iterated the 
probabilistic decoding algorithm \cite{MacKay}, at different 
values of the noise channel $p$, for an ensemble of $10^3$ networks 
of $10^4$ and $2 \times 10^4$ nodes respectively. The inset shows 
the calculated mean values, the corresponding error being smaller than the
corresponding symbol size. A precise estimate of the dynamical
transition point has been obtained considering the percentage of
decoding success within the network ensemble as a function of the
noise level, for different system sizes, \cite{KS}. The decoding
experiments were performed for different choices of the
temperature. Along the Nishimori line (see 
Fig.~\ref{ldpc_phase}) we find the value of $p_d \simeq 0.1665 \pm 
0.0005$ in excellent agreement with the reported value for regular
$R=1/4$ Gallager codes~\cite{RU}.

Notice that the reentrant nature of both the dynamical and critical
boundaries, as we found solving the convolution (\ref{crux2}), is
confirmed by the decoding experiments. In the range of noise
levels $0.1315<p<0.1665$, decoding is possible only in a range of
finite temperatures , as indicated in the inset of
Fig.~\ref{ldpc_phase}. The boundary between decodeable and
undecodeable regions reaches its maximal value at the Nishimori
temperature; zero temperature decoding fails at the same noise
level. The reentrant nature of the critical phase boundary
increases below the Nishimori line, but eventually weakens as the
temperature is decreases further, a phenomenon that has already
been discussed in a variety of different
contexts~\cite{MKKB,Hartmann}.

The fact that MPM and zero temperature BP/RS decoding do not
coincide is not in contradiction with the results obtained
in~\cite{vMSK} which indicate that MAP and MPM
decoding provide the same critical transition point. Clearly, zero
temperature BP/RS decoding does not provide MAP results due to RSB
effects. The latter and their effect on the phase diagram and on
the reentrant nature of the phase boundaries will be further
discussed in section~\ref{sec:RSB}.

An independent calculation has been performed at zero temperature. 
In this case the recursion relations (\ref{recursion}) simplify 
and one finds limiting expressions for the two functions $u( \{ x_i\},J)$,
$a(\{x_i\},J)$. Implementing numerically such asymptotic expressions
would require an independent calculation where one explicitly changes the
recursion relations (\ref{recursion}) to their asymptotic
$\beta \rightarrow \infty$ form. 
The importance of the polynomials $R(\sigma )$ becomes clear at
this point.
At any temperature considered, we distinguish the finite and zero temperature 
contributions, corresponding to factorize the largest exponent in the 
sum (\ref{poly}), so that the limit of zero temperature is particularly
simple to handle in our approach and no explicit changes are
required in the recursion relations (\ref{recursion}). We observe that
at zero temperature the local field distribution is not peaked on
integers as one might expect.
We find that the relative weight of the background behind the integer-peaked
distribution weakens as one consider the RSB solution, but
restricting the distribution of the local fields to integer 
values, which results in a modification of the recursion relations
(\ref{recursion}) leads to different results. For LDPC (RS)
decoding in the BSC this leads to an erroneous dynamical
transition value of $p_d \simeq 0.121$, marked in
Fig.~\ref{ldpc_phase} by a black filled dot, well below the
reported zero temperature value of $p_d \simeq 0.1315$. Similar
behavior occurs in the binary erasure channel problem if one
simply neglects the background distribution, as we will discuss in
section~\ref{sec:RSB}.

\begin{figure*}
\centering
\includegraphics*[scale=0.6]{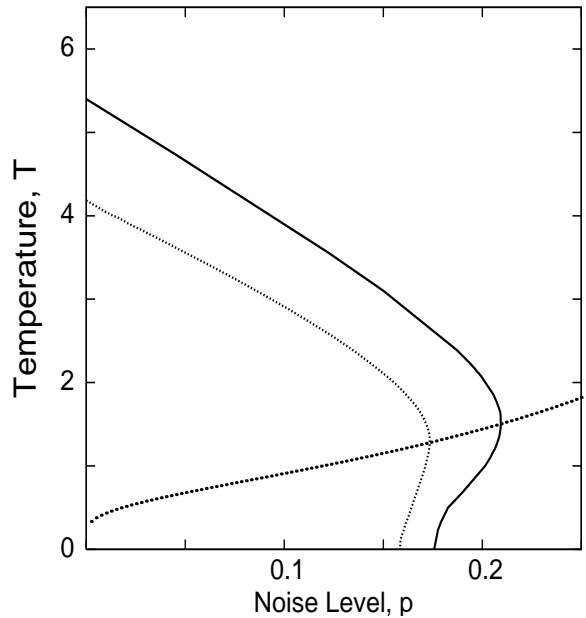}
\caption{The phase diagram of LDPC decoding for the BSC with the
initial condition (\ref{ic}). The location of the critical
transition in unchanged unlike the location of the dynamical
transition, that has been observed to increase. We observe, on the 
Nishimori line, a dynamical transition occurring at $p_d \simeq 0.1734$, 
well above the ordinary dynamical transition of the regular 
Gallager code of rate $R=1/4$.
\label{initial_cond}}
\end{figure*}

The initial condition $\pi^0(x)=(1-q) \delta(x-1)+q
\delta(x+1)$, at $q=\frac{1}{2}$, has been used as 
initial condition for the BSC analysis.
A dependence of the solution of
the convolution (\ref{crux3}) on the initial condition for the local
field has been observed. 
We have studied the above convolution (\ref{crux3}) for
different choices of the initial condition in the form 
\begin{equation}
\pi(x)=(1-q)~ \delta(x-1)+q~ \delta(x+1).
\label{ic}
\end{equation}
The choice of $q=p$, where $p$ is the flip rate of the channel
results in the phase diagram of  Fig.~\ref{initial_cond}. {\em Any}
choice of initial condition with $q>p$ (including the case $q=1/2$
considered above) leads to the dynamical transition line (dashed
line in Fig.~\ref{ldpc_phase}), except for the choice $q=p$ which
leads to a different dynamical transition. The location of the
critical transition remains unchanged.

For instance, using the initial condition $q=p$, the dynamical
transition along the Nishimori line is increased to $p_d \simeq 0.1734$,
well above the known dynamical transition point.
The physical interpretation of this behavior remains to be explained.
Since it results from a simple change in the initial conditions 
of (\ref{crux}), one might argue that a properly modified decoding 
algorithm would enable us to decode above current  $p_d$ values; this is a 
possibility that we cannot rule out and further research is needed in 
order to elucidate this point.

We where able to perform a second set
of decoding experiments that coincide with the dynamical
transition line in Fig.~\ref{initial_cond} but only for noise level 
values below the value $p_d \simeq 0.1665$. 
The reentrant nature of the dynamical transition
line (the dynamical transition at zero temperature being 
$p_d \simeq 0.158$) is reduced in this case.

\subsection{Irregular Gallager Codes}
\label{sec:Irregular}
For a constant $K$ value one can select non-integer connectivity
values of average $\bar{C}$ and consider a modified form of the
convolution (\ref{crux3}) that involves a linear combination of
terms corresponding to different connectivity values. Non-integer
values of the multi-spin interaction parameter $K$ can also be
easily accommodated in (\ref{crux3}).

It has been reported that irregular low-density parity-check codes 
outperform regular constructions~\cite{MacKay,KS}. 

\begin{figure*}
\centering
\includegraphics*[scale=0.6]{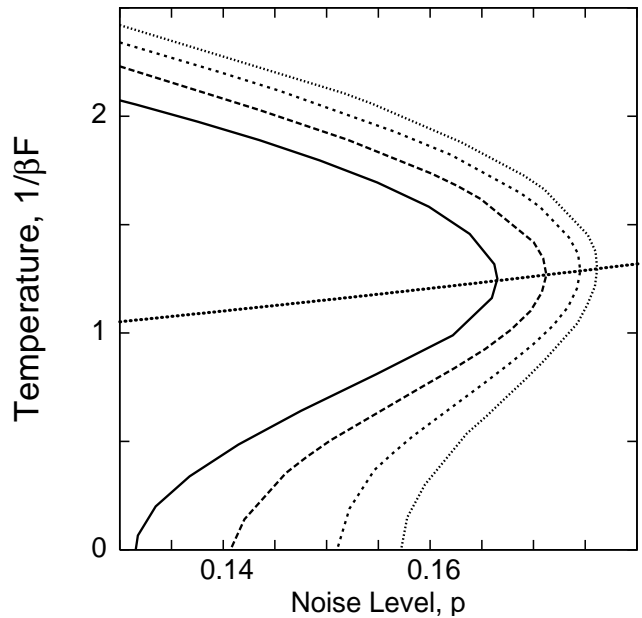}
\caption{ Irregular LDPC phase diagram for the code rate R=1/4 . We show
 the dynamical transition for different values of the connectivity
 parameters $C$ and $K$. The solid line corresponds to the dynamical 
 transition line of a regular construction with $C=3$ and $K=4$. 
 The dashed, dotted and thin-dotted lines correspond respectively to
 $\bar{C}=2.9,2.8,2.7$, at constant rate $R=1/4$.
 At lower average connectivity values, e.g. along
 the Nishimori line, the convolution (\ref{crux3}) no longer flows
 to a ferromagnetic sink, but we observe a stable crossover to a
 distribution with large but finite average, so that successful decoding
 cannot be expected.
\label{irregular}}
\end{figure*}

We considered a generalized form of the convolution (\ref{crux3}), (an
explicit form is not provided for brevity) where both connectivity
$C$ and multi-spin interaction $K$ take non-integer values such
that the rate $R=1-\bar{C}/ \bar{K}=1/4$ is preserved.
In Fig.~\ref{irregular} we report the dynamical transition line
for the regular construction $C=3,K=4$ ($R=1/4$) already considered in the
previous section together with the dynamical transition lines for
irregular constructions of two values $K=4$ and $K=3$ and 
$C=2$ and $C=3$, such that 
$\bar{C}=2.9,2.8,2.7$ (left to right) and $\bar{K} \simeq 3.87,3.74,3.6$.
The dynamical transition value is observed to increase for the irregular constructions.
This does not give a direct information on the
specific pattern of irregularity that one should consider 
but provides quantitative information on the average
connectivity/multi-spin interaction values one should investigate
further. As for the regular case discussed earlier, a phase
diagrammatic approach to the problem of irregular LDPC codes 
is important to understand the reason behind the successful
decoding properties of irregular LDPC codes. A
similar behavior to the one shown in Fig.~\ref{irregular}, but less
pronounced, has also been observed for LDPC decoding with the initial
conditions (\ref{ic}) when $q=p$.
\subsection{Finite connectivity problem with $C/K>1$}
\begin{figure*}
\centering
\includegraphics*[scale=0.5]{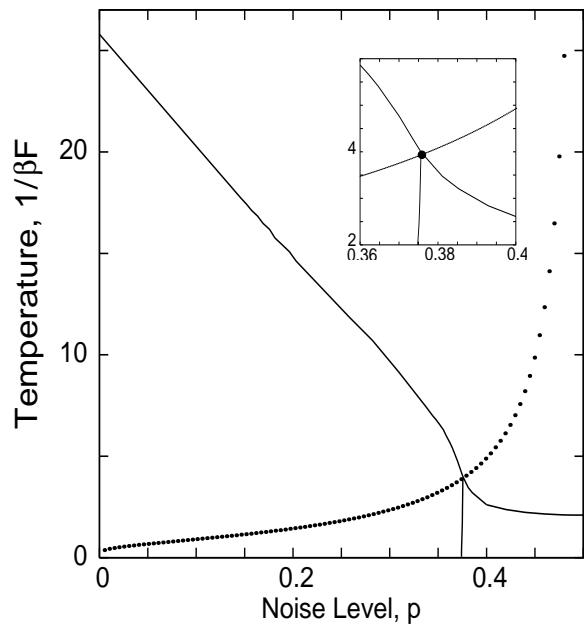}
\caption{ The dynamical transition line for the finite-connectivity
system with $C=9$, $K=4$.
We report the location of the dynamical transition, shown now in
thick solid lines. At high enough values of the flip rate, 
the convolution (\ref{crux3}) either flows to a
stable, asymmetric local field distribution (at high enough
temperatures), or to a symmetric SG one (below the
 Nishimori line). Before flowing to the
corresponding sink, the flows of the convolution (\ref{crux3}) crosses 
over in the vicinity of either an unstable distribution above
the Nishimori line, shown as the dotted line, or in the vicinity of a
strong coupling unstable local field distribution below the Nishimori
 line. Similar behavior has been observed for the LDPC convolution
(\ref{crux3}) with the ordinary initial condition $q=1/2$. In that case
the location of the dynamical transition occurs at lower values of the
noise channel probability. The inset shows the location of the dynamical
boundaries together with the Nishimori line.
\label{c9k4}}
\end{figure*}

We considered the finite-connectivity model (\ref{RFRB}) for the 
values of the connectivity $C=9$ and $K=4$. Despite the fact that it does not 
correspond to any Gallager LDPC code construction it is interesting to 
consider these connectivity parameters for two reasons:
1) They can represent code constructions in other variants of the LDPC
family such as Sourlas or MN codes \cite{MacKay}, and as such may also 
be linked to other LDPC-based applications.
2) They give rise to a SG behavior which is of interest. 
In Fig.~\ref{c9k4} we report the corresponding phase
diagram obtained considering the convolution (\ref{crux3}).
An SG phase occurs in this case. At low values of the flip rate $p$ the 
convolution (\ref{crux}) flows to the ferromagnetic sink. In the 
vicinity of the dynamical transition line (marked in Fig~\ref{c9k4} as
the solid black line) the trajectories under iteration of the above 
convolution cross over to an unstable distribution and then either flow to the
ferromagnetic sink, below the dynamical transition, or to a stable 
non-ferromagnetic distribution. For temperatures below the Nishimori 
line, at high enough values of the flip rate, the convolution instead 
flows to a spin-glass distribution.

\section{Replica Symmetry Breaking}
\label{sec:RSB}
The role of RSB for the problem (\ref{RFRB}) is the subject of
this and the following sections. The one-step RSB solution for the
Bethe SG, as a special limiting case of the Hamiltonian
(\ref{RFRB}), has been studied in detail long ago. In this case it
is well known that the RS solution is incorrect~\cite{MP2}.
Recently, the onset of RSB phenomena in some limiting cases of
the problem (\ref{RFRB}) has been considered, relying on the
variational approximation~\cite{Mon} (for a recent formulation see
\cite{Kabashima}). Consider the equation for the global order
parameter $g(\{ \sigma^{ \alpha } \})$, firstly introduced for the 
Bethe SG by Mottishaw,~\cite{Mo} in the general case of a
multi-spin interaction parameter $K$,
\begin{eqnarray}
g(\{ \sigma_o^{ \alpha } \}) &=& E_{ J, \zeta }\big [ Tr_{ \{
\sigma_i^{ \alpha} \} }
 g^{C-1}( \{ \sigma_1^{ \alpha } \}) \cdots
g^{C-1}( \{ \sigma_{K-1}^{ \alpha }\} ) \nonumber \\
&& \times \exp [ \beta J \sum_{ \alpha} \sigma_o^{ \alpha}
\sigma_1^{ \alpha} \cdots \sigma_{K-1}^{ \alpha} +  \beta F \sum_{ j
\alpha} \sigma_j^{ \alpha}  ] \big ]. \nonumber \\
\label{gopK}
\end{eqnarray}
where $ \alpha=1,\ldots, n$ labels different replicas of the system.
The above expression holds in general, whether replica symmetry
is broken or not (see Appendix B for the derivation).
In the RS approximation one assumes that different
replicas are equivalent and the functional order
parameter $g(\{ \sigma^{ \alpha } \})$ depends only on the variable
$ \sigma = \sum_{ \alpha } \sigma^{ \alpha}$. In the limit of
$n \rightarrow 0$, can be shown to satisfy,
\begin{eqnarray}
g( \sigma ) = E_{ J, \zeta} \int \prod_{j=1}^{K-1} \frac{ ds_j}{2 \pi}
g^{C-1}( i s_j)~
du_j~ e^{ i u_j s_j} \times \nonumber \\
\exp [ - \sigma / \beta \tanh^{-1} ( \tanh( \beta
             J) \prod_{j=1}^{K-1} \tanh( \beta u_j - F \zeta) )
             ]. \nonumber \\
\label{RS}
\end{eqnarray}
This equation corresponds to the equation that have been
written for the Bethe SG, as soon as $K=2$ and for vanishing message prior
temperature $F=0$~\cite{Mo}. A derivation of equation (\ref{RS})
is found in Appendix B, where we show the equivalence of the above
expression with the set of recursion relations (\ref{crux}), for
the local field and conjugate distributions. In the one-step RSB
approximation one assumes instead that the functional order
parameter $g(\{ \sigma^{ \alpha } \})$ depends only on the
variables $ \sigma_M= \sum_{ \gamma} \sigma_{M_{ \gamma}}$, where
$\alpha=(M, \gamma)$ is the replica index, $M=1, \ldots ,n/m$ and
$\gamma= 1, \ldots ,m$. In this case one finds that the
corresponding equation for the functional order parameter $g(\{
\sigma^{ \alpha } \})$ becomes
\begin{eqnarray}
g(\{ \sigma_M \}) = { \cal N}^{-1} E_{J, \zeta} \int
\prod_{j=1}^{K-1} \prod_{M=1}^{n/m}
 \frac{ds_{Mj}}{2 \pi} g^{C-1}( i s_{Mj} ) \nonumber \\
\int \prod_{Mj} dr_{Mj} e^{i r_{Mj}s_{Mj}}
\exp \Big [ - \sum_M \sigma_M u (\{ r_{Mj} \},J) \Big ] \times \nonumber \\
\exp \Big [ \frac{m}{2} \sum_M a( \{ r_{Mj} \},J ) \Big ],\nonumber
\label{RSB}
\end{eqnarray}
where the normalization is given by
\begin{eqnarray}
{ \cal N} =  \int \prod_{j=1}^{K-1} \prod_{M=1}^{n/m} \frac {ds_{Mj}}{2 \pi} g^{C-1}( i s_{Mj} ) \nonumber \\
\int \prod_{Mj} dr_{Mj} e^{i r_{Mj}s_{Mj}} \prod_j \cosh^{m} ( \beta r_{Mj}).
\end{eqnarray}
The corresponding one-step RSB convolution
follows, assuming that the order parameter factorize according to
\begin{equation}
 g(\{ \sigma_M \}) = \prod_{M=1}^{n/m} f( \sigma_M).
\label{factorized}
\end{equation}
The assumptions behind the {\em factorized} ans\"atz
(\ref{factorized}), where the local field distribution is independent 
from the intra-group index, are discussed in~\cite{GL}. 
In this case the one-step RSB convolution (\ref{RSB}) become
\begin{eqnarray}
\pi(x)&=& E_{ \zeta} \int \prod_{i=1}^{C-1} d \hat{x}_i \hat{ \pi
}(\hat{x}
 ) \delta \Big ( x - \sum_{i=1}^{C-1} \hat{x_i}- F \zeta \Big) \nonumber \\
\hat{ \pi }( \hat{x} ) &=& { \cal N} ^{-1} E_J \int \prod_{j=1}^{K-1}
dx_j \pi(x_j) \exp [ \mu a( \{ x_j\},J) ]  \times \nonumber \\
&&\delta \Big( \hat{x}-u( \{ x_j\},J) \Big), 
\label{crux_rsb}
\end{eqnarray}
where the normalization ${ \cal N}$ is given by
\begin{equation}
{ \cal N} =  \int \prod_{j=1}^{K-1}
dx_j \pi(x_j) \exp [ \mu a( \{ x_j\},J) ],
\end{equation}
and where $\mu = m/ \beta$. The above equations are valid, under
the assumption (\ref{factorized}), at finite temperatures and
arbitrary connectivity values $C$ and $K$. After its very first
formulation~\cite{GL} the convolution (\ref{crux_rsb}) has been
studied in a range of different contexts, from random
$K$-SAT to the coloring problem~\cite{FLZ}. Using the identity,
\begin{equation}
\exp [ \mu a( \{ x_j\},J)] = \Big [ \frac{ 2 \cosh ( u( \{ x_j\},J) )}{ \prod_j \cosh ( \beta x_j ) } \Big ]^{ \mu},
\label{identity}
\end{equation}
and substituting it in equation (\ref{crux_rsb}) one recovers the
one-step RSB equations that have been also derived using the
variational approach~\cite{Mo,Franz,Kabashima}. This shows that
the so called frozen variational ans\"atz, considered in
\cite{Mon} and the factorized ans\"atz of~\cite{GL} are
equivalent. Even though there exists
some degree of arbitrariness~\cite{MZ} in the way one
redistributes the re-weighting between the two equations
(\ref{crux_rsb}), we prefer the notation that involves the
exponential of the energy term $a(\{ x_j\})$, rather than the less obvious (and
numerically more difficult to implement) parallel expression 
that follows considering the right hand side of
(\ref{identity})~\cite{Kabashima,FLZ}. The simplified expression
of the free-energy presented in section III was obtained in virtue
of the same identity (\ref{identity}), and the same degree of
freedom writing the free-energy was already discussed for the
Bethe SG~\cite{MP2} (the right hand side of (\ref{identity})
being, at $K=2$ the function $c(x,J)$ discussed in ref. \cite{MP2}).

\subsection{The Bethe SG}
\label{sec:Bethe}
\begin{figure*}
\centering
\includegraphics*[scale=0.6]{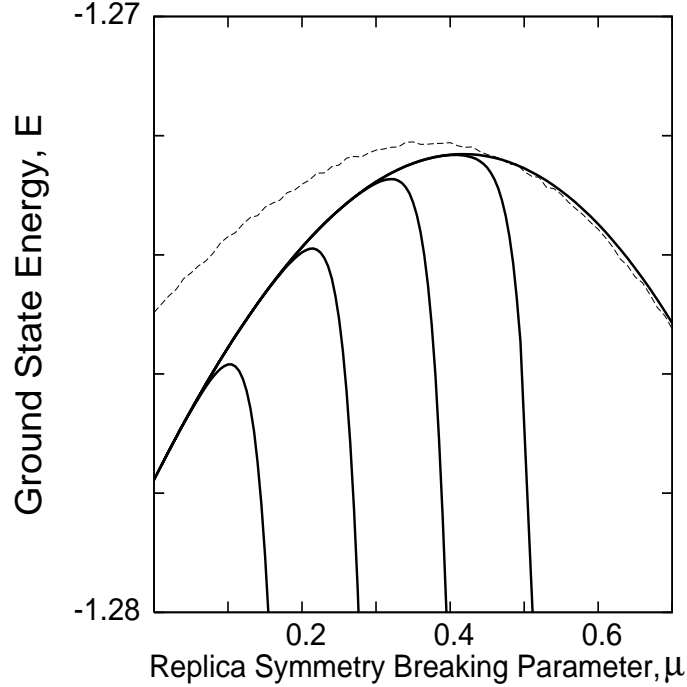}
\caption{ The ground state energy as a function of the replica symmetry
 breaking parameter for different values (from left to right) of the bias
 $\rho=0.66,0.68,0.70,0.72,0.74$. We also show in thin dotted line the ground
 state energy including the background distribution as in ref.\cite{WS}.
\label{Bethe_GSE}}
\end{figure*}

\begin{figure*}
\centering
\includegraphics*[scale=0.6]{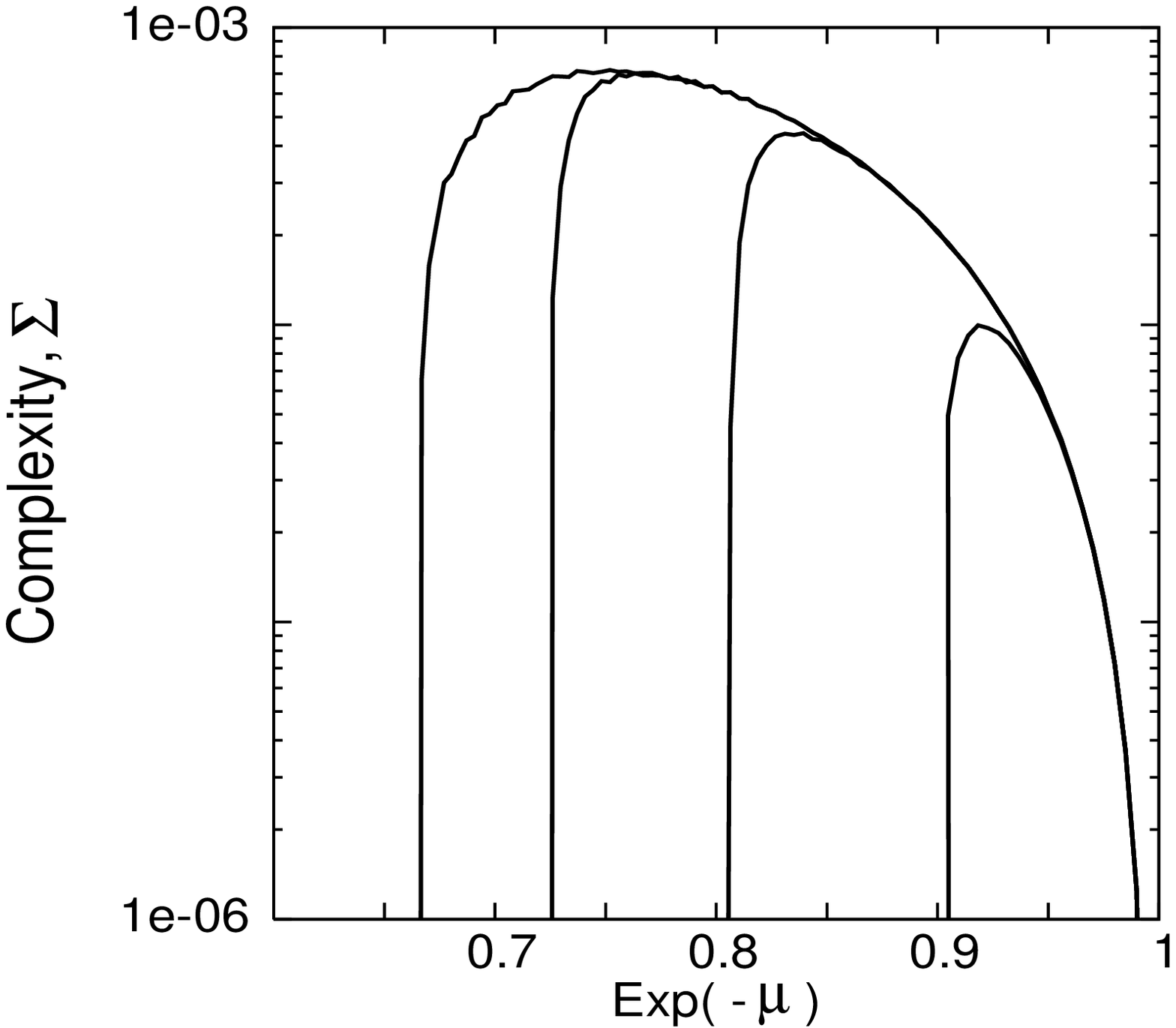}
\caption{ The complexity $ \Sigma$ as a function of $ \exp (- \mu)$ for
 different values of the ferromagnetic bias $
 \rho=0.66,0.68,0.70,0.72,0.74$, within the three peak ans\"atz. 
\label{Bethe_Compl}}
\end{figure*}

\begin{figure*}
\centering
\includegraphics*[scale=0.6]{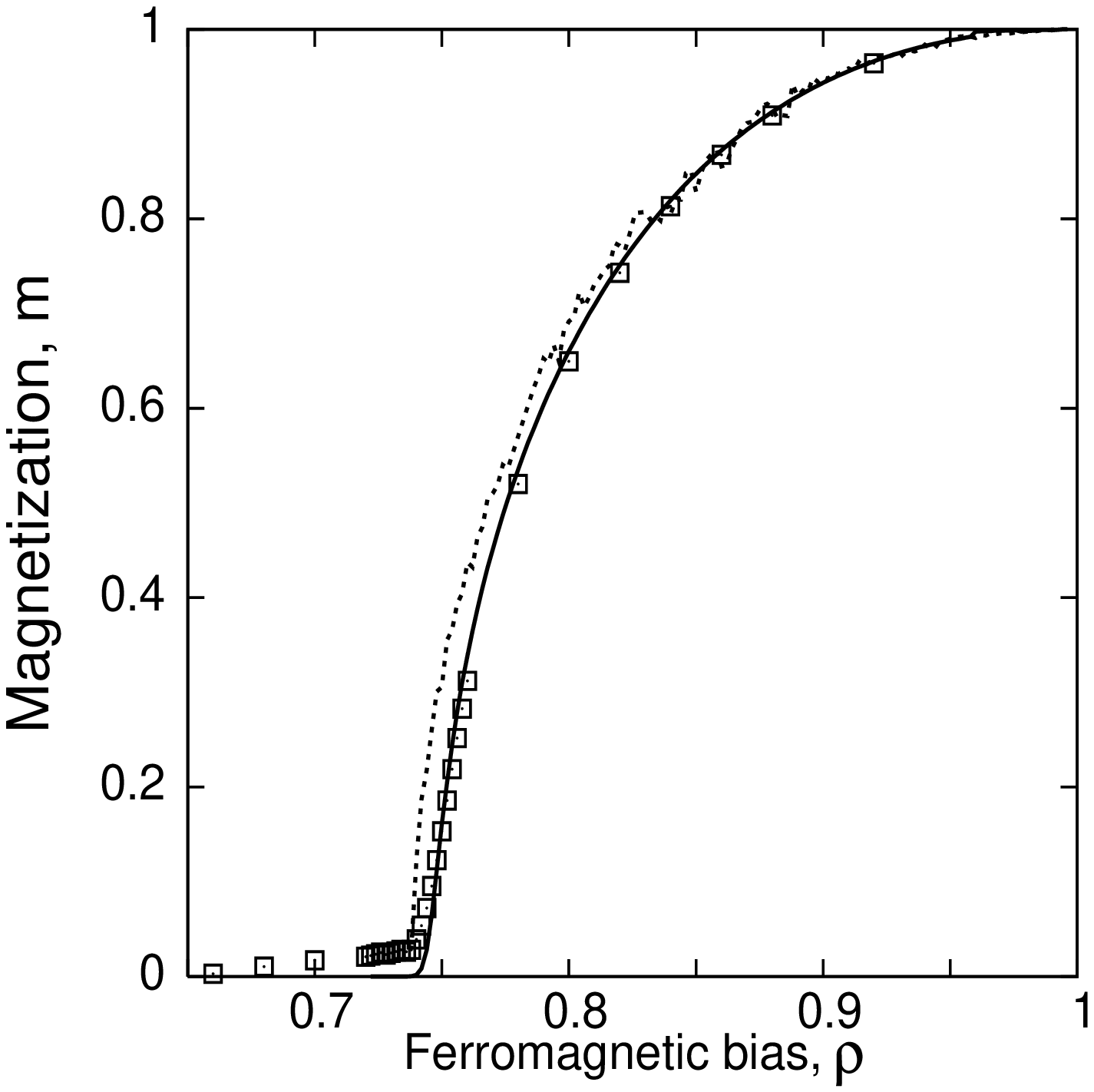}
\caption{ The magnetization as a function of the ferromagnetic bias $
 \rho $. The thick solid line correspond to the RS calculation obtained with
 $4096 \times 4$ bins. Square boxes correspond to the one step RSB calculation
if one neglects the background local field distribution. Finally the
 dotted line represents the one step RSB calculation including the
 background calculation, according to equation (\ref{crux_rsb}) and
 was obtained with $32768 \times 4$ bins. 
\label{Bethe_Mag}}
\end{figure*}

We apply the method described in section III to study the one-step
RSB convolution (\ref{crux_rsb}) of the Bethe SG, which has been
extensively studied and recently considered also in its
ferromagnetically biased version~\cite{Castellani}. In this case the
distribution of the quenched random variable $J$ is
\begin{equation}
P(J)= \frac{1 + \rho}{2} \delta(J-1)+ \frac{1 - \rho}{2} \delta(J+1),
\end{equation}
and where the parameter $ \rho \in [0,1]$ is $ \rho=1$ for a
ferromagnet and $ \rho=0$ for the unbiased Bethe SG. The Bethe SG
corresponds indeed to the original problem (\ref{H_0}) at $K=2$.
We will present results obtained at
zero temperature. In the zero temperature limit we find the
following expression for the free energy $\Phi( \mu)$,
\begin{equation}
\Phi(\mu)= \Delta E^{(1)} - \frac{C}{2} \Delta E^{(2)},
\end{equation}
where explicit expressions for the site and link contributions to
the ground state energy are
\begin{eqnarray}
\exp \Big ( && \hspace*{-4mm}- \mu \Delta E^{(1)} \Big) = \int \prod_{j=1}^C
 dx_{j} \pi(x_{j}) \times \nonumber \\
&& \exp \Big ( \mu \sum_{j=1}^C a( \{ x_j\},J_j) + \mu 
| \sum_{j=1}^C u( \{ x_j\},J_j) | \Big ) \nonumber \\
\exp \Big ( &&  \hspace*{-4mm}- \mu \Delta E^{(2)} \Big) = \int \prod_{j=1}^2
 dx_{j} \pi(x_{j}) \times \nonumber \\
&& \exp \Big ( \mu \max_{ \sigma_1, \sigma_2} ( x_1 \sigma_1
 +x_2 \sigma_2 + J \sigma_1 \sigma_2) \Big ), \nonumber \\
\end{eqnarray}
corresponding, at $ \mu=0$ and $K=2$, to the RS equation
(\ref{free-energy}) in the zero temperature limit and
neglecting the prior term in (\ref{RFRB}). Similar expression have 
been discussed in~\cite{MP2}. In the limit of zero temperature one finds
$a(J,x)= |x|+ \delta_{x,0}$ and $u(x,J)=J S(x)$, being
$S(x=0)=0,$ and $S(x\ne0)= sgn(x)$. In what follows we 
present results obtained under the $integer~peaked$ ans\"atz as
well as for the full distribution. The relative importance of
non-integer values in the distribution that solves the above
convolution (\ref{crux_rsb}) is rather limited in this case. The
background behind the integer peaked values weakens as one
consider the one-step RSB case, and may disappear in
a fully replica symmetry broken scenario~\cite{MP2}. In Fig.~\ref{Bethe_GSE} we
show the ground state energy as a function of the replica symmetry
breaking parameter $ \mu$ for different values of the
ferromagnetic bias using the integer peaked ans\"atz and in the
case of the full local field distribution. The ground state energy
one extrapolates including the background calculation is higher
than the one obtained within the three peak ans\"atz~\cite{GL}.
In Fig.~\ref{Bethe_Compl} we show the corresponding complexity or configurational
entropy, that can be obtained as a parametric plot in terms of 
the RSB parameter $ \mu $, the free-energy $f(\mu)$
being its Legendre transform~\cite{MP1}. We observe a positive
complexity for all reported values of $\rho$. At higher values of
the ferromagnetic bias $\rho>\rho^* \simeq 0.74$ a positive
complexity is not observed anymore. One should also consider that
the above values should be extrapolated in the limit of large
number of bins, so that one expects a slightly higher value $ \rho^* \simeq 0.75$.
In Fig.~\ref{Bethe_Mag} we show the magnetization for in the replica symmetric
and one-step replica symmetry broken solution at different values
of the ferromagnetic bias. At values $ \rho < 0.65$ one recovers
the results known for the unbiased Bethe SG.  The solid line
indicates the magnetization computed in the replica symmetric
approximation (notice that this is a computed magnetization for a
finite number of bins, so that it approaches the expected value of
$\rho_{RS}=0.75$ from below), while the square boxes represent the
magnetization in the one step RSB approximation without
considering the background. For each value of $ \rho$, we
extrapolated the magnetization that corresponds to the value of $
\mu$ where the ground state energy has a maximum. Similarly, the 
dashed line points to the results obtained considering the true
convolution (\ref{crux}), that includes non integer values of the
local field. The putative size of the so called $mixed~phase$,
previously discussed~\cite{Castellani}, if any, is rather small
and do not agree with previous estimates (we observe the
magnetization within the RSB calculation to vanish at values of $
\rho \simeq 0.75$, if one considers a large enough number of bins
within the calculation, implying that the size of the mixed phase,
if any, is vanishingly small). It should be said however that the
calculation of~\cite{Castellani} considered a full one step RSB
calculation, in the sense of~\cite{MP2}, where the factorized
approximation is relaxed. To investigate this point better, we
performed the following: instead of implementing the convolution
(\ref{crux}) within the factorized approximation~\cite{GL}, we
choose a population of $5 \times 10^5$ distributions (including a
site dependence). At each step of the iteration process we
consider pairwise-convolutions of local field distributions
corresponding to different sites, selected at random; after a
transient regime, we obtain the equilibrium population. Waiting a 
number of iteration steps of the order of the population size,
according to our calculation, the local field distributions become
site independent, meaning that the factorized approximation
provides similar results to those of the full RSB ans\"atz. If one
compares Fig.~\ref{Bethe_Compl} with the corresponding plot in ref.
\cite{Castellani}, where a similar calculation of the complexity
is presented, we see that the location of the maximum of the
complexity as a function of $ \mu$ {\em decreases} for increasing
$ \rho$ values (see Fig.~\ref{Bethe_Compl}), as one might 
intuitively expect, increasing the ferromagnetic bias.

\subsection{The Binary Erasure Channel}
\label{sec:BEC}
\begin{figure*}
\centering
\includegraphics*[scale=0.7]{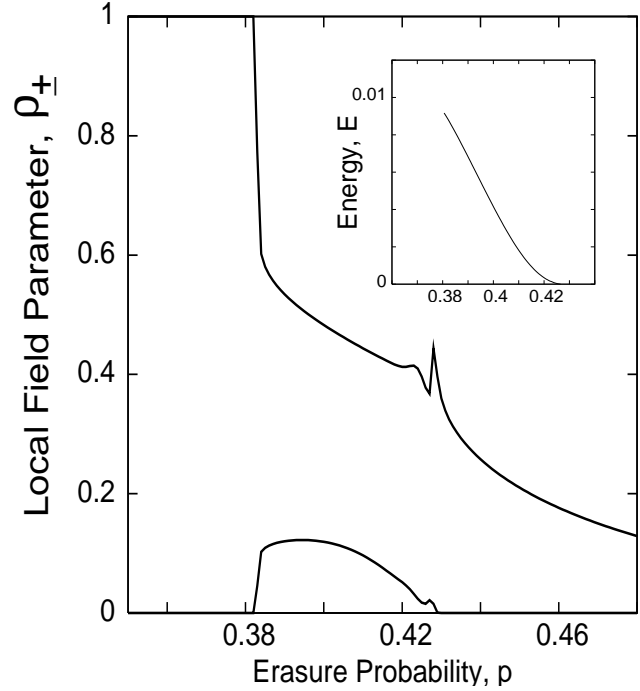}
\caption{For the code rate R=1/2 and a BEC channel we report the local
 field parameter $ \rho_{ \pm}$ as a
function of the erasure probability $p$, at zero temperature, within the
integer value ans\"atz (\ref{three_peaked_ansatz}). The continuous lines
represent respectively the probability of a negative $ \rho_{-}$ and
positive $ \rho_{+}$ local field. The inset shows the computed energy as
a function of the erasure probability. 
\label{bec_1}}
\end{figure*}

\begin{figure*}
\centering
\includegraphics*[scale=0.7]{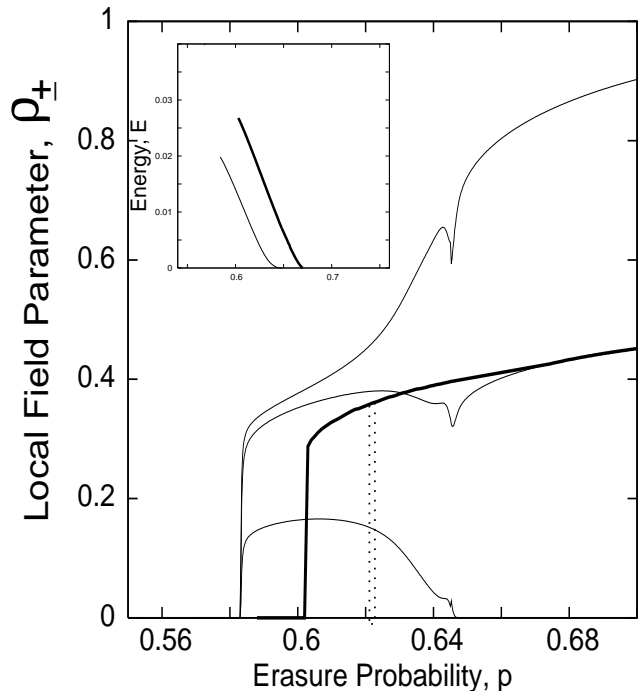}
\caption{ For a code rate R=1/4 and a BEC channel at zero temperature, we
report the local field parameters as a function of the erasure probability
 $p$, for the ans\"atz (\ref{three_peaked_ansatz}) shown in thin lines,
corresponding to $\rho_{-}$,$~  \rho_{-}+ \rho_{0}/2$ and $ \rho_{0}$.
We then report (thick line) the results for the local field probability
according to the full convolution (\ref{crux_bec}), as it is in the
simplest RS approximation $ \mu=0$,
corresponding to $\rho_{x<0}$, being $\rho_{x>0}=1-\rho_{x<0}$. The
dashed vertical line corresponds instead to the location of the dynamical
transition within the one step RSB calculation.
The inset shows the computed energy as a function of the erasure probability.
within the {\em three peak} ans\"atz (thin line) and in the case of the 
full convolution of the local field being considered (thick solid line).
\label{bec_2}}
\end{figure*}

In the case of the BEC discussed in section II,
the form of the convolution (\ref{crux3}) simplifies and the
distribution of the local field, and its conjugated distribution
become bimodal. Despite its simplicity the binary
erasure channel problem is of practical relevance~\cite{LMSS,RU2}.
For the BEC one writes the convolution
(\ref{crux3}) as
\begin{eqnarray}
\pi(x)=\rho \delta(x)+(1-\rho) \delta_{ \infty}(x) \nonumber \\
\hat{ \pi}( \hat{x} )=\hat{ \rho} \delta( \hat{x} )+(1-\hat{ \rho})
 \delta_{ \infty}( \hat{x}) \nonumber \\
\end{eqnarray}
where $\delta_{ \infty}( \cdot )$ is a delta function located at
$+ \infty$, $\rho(x)$ is the distribution of the effective local
field and $ \hat{ \rho}( \hat{x})$ its conjugate distribution. The
resulting convolution~\cite{Franz} involving the two
distributions $\rho(x)$ and $ \hat{ \rho}( \hat{x})$, in the
one-step RSB case, is given by
\begin{eqnarray}
\rho(x) &=& \int \prod_{l=1}^{C-1} d \hat{x}~ \hat{ \rho} (\hat{x} )~ \delta
 \Big ( x - \sum_{l=1}^{C-1} \hat{x}_l \Big ), \nonumber \\
\hat{ \rho}( \hat{x}) &=& \sum_{ \nu}^{K-1} f_{ \nu} { \cal N}_{ \nu}
\int \prod_{i=1}^{ \nu }
dx_i ~\rho(x_i)~ \exp [ \mu a( \{x_i \},J) ] \times \nonumber \\
&& \delta \big ( \hat{x}-u(\{ x_i\},J) \big ), \nonumber \\
\label{crux_bec}
\end{eqnarray}
where $f_{ \nu}$ are the binomial coefficients of order $K-1$ and ${\cal
N}_{ \nu}$ the set of normalization coefficients defined by
\begin{equation}
{\cal N}_{ \nu} = \int \prod_{i=1}^{ \nu} dx_i \rho(x_i) \exp [ \mu a( \{x_i \},J)].
\end{equation}
One can readily check that the above convolution reduces to the
variational approximation~\cite{Mon,Franz} using 
identity (\ref{identity}). We note that 
the variational ans\"atz, as already discussed for the case of the
BSC, coincides with the factorized ans\"atz
(\ref{factorized})~\cite{Mo,GL}. We consider an explicit
expression for the one step RSB free-energy that involves the
binomial coefficients of order $K$. Within the factorized
hypothesis~\cite{GL}, the expression for the free-energy is easily
computed as a function of the RSB parameter $ \mu$. We present
results for the BEC at zero temperature for
values of the code rate R=1/4 and $R=1/2$ (note that we also 
consider in what follows the integer peak ans\"atz for the local
field distribution~\cite{Franz} in order to compare with the
general case of non-integer local fields). In the inset of Fig.~\ref{bec_1}
we show the energy as a function of the erasure probability $p$
for the rate value $R=1/2$ within the integer valued local field
approximation. Notice that the integer value approximation
corresponds to having only three bins in our approach of section
III. In Fig.~\ref{bec_2} we compare the results within and without the
integer approximation for a code rate $R=1/4$. The dynamical
transition is defined as the point where sub-optimal solutions emerge 
the corresponding complexity being non zero, as already discussed for
the BSC channel (see Fig.~\ref{bec_2}). The location of the dynamical 
transition is also the point where the recursion (\ref{crux_rsb}) flows to a
subdominant distribution, rather than the ferromagnetic sink.
This coincides with the point where algorithms devised for real
decoding experiments fail to decode the original message. Within
the RS framework, we find that the location of the dynamical
transition changes significantly if non-integer values of the
local field are taken into account. Within the three peak ans\"atz
the local field distribution is simply defined as~\cite{Vicente}
\begin{equation}
\rho(x) = \rho_{+} \delta(x-1) + \rho_{-} \delta(x+1) + \rho_0 \delta(x)
\label{three_peaked_ansatz}
\end{equation}
The calculation we performed within the integer local field
approximation (for rates $R=1/2$ and $R=1/4 $ in Fig. ~\ref{bec_1} and 
Fig.~\ref{bec_2} respectively) might suggest to interpret the point 
$p^{*} \simeq0.4294$ for the
rate value $R=1/2$ as the dynamical transition point~\cite{Franz}.
If however one compares (see Fig.~\ref{bec_2}) with the results of the
local field distribution that now includes non-integer values (the
thick solid line corresponding to $\rho_{x>0}$) we see that the location 
of the dynamical transition is rather different. Moreover the behavior of the
local field probability $\rho_{ \pm }$ around the value $p^{*}
\simeq 0.4294 $, within the $three~peak$ ans\"atz is rather
suspicious. Indeed, including non-integer values of the
local field turns out to be crucial for having a quantitative
description of the dynamical transition and possibly explains the
disagreement with numerical experiments that has been reported
lately for the R=1/2 BEC problem \cite{Franz}. If one avoids the
{\em three peak} ans\"atz, the nature of the flows dictated by the
recursion (\ref{crux_bec}) is smooth as it always leads
to a well-behaved behavior of the probability densities $\rho_{x
> 0}$ and $ \rho_{ x < 0}$. Up to this point all considerations 
are still within the RS approximation.

\begin{figure*}
\centering
\includegraphics*[scale=0.6]{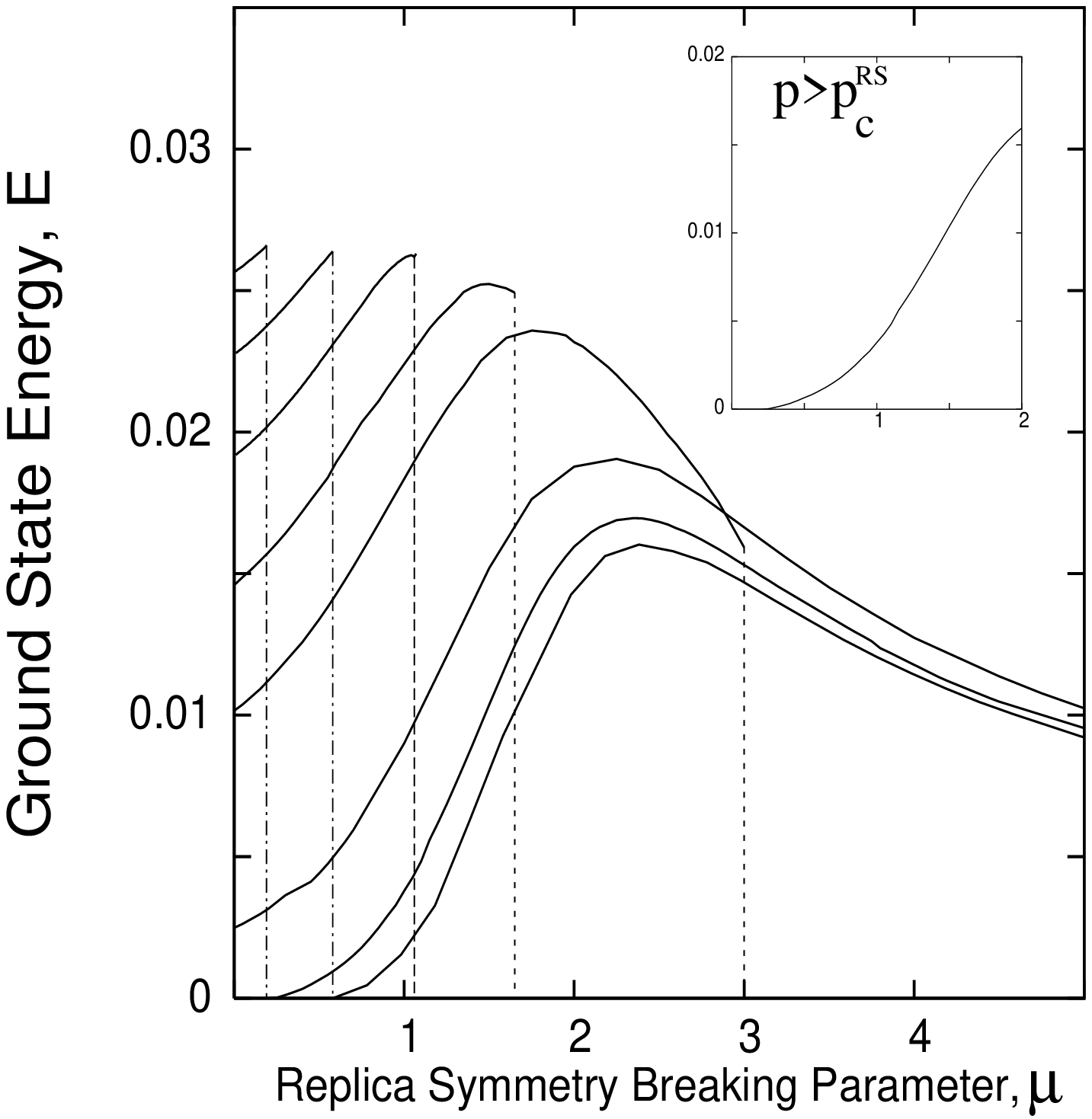}
\caption{The one step RSB free energy as a
function of the parameter $ \mu$ for different values (left to right) of 
the erasure probability $p=0.606,0.613,p=0.62,0.63,0.64,0.66$, and
$p=0.67$. The first two curves, corresponding to y $p=0.606$,$~p=0.613$ 
drop to zero before a maximum is reached as a function of $\mu$,
indicating the absence of the physical branch of the
complexity. Differently, around values of $p \simeq 0.625$ the energy 
reaches a maximum { \em before} dropping to zero, so that the physical 
branch of the complexity appears. The inset shows the ground state
energy as a function of the RSB parameter $ \mu$, at a flip rate above
the critical transition point.
\label{bec_3}}
\end{figure*}

\begin{figure*}
\centering
\includegraphics*[scale=0.6]{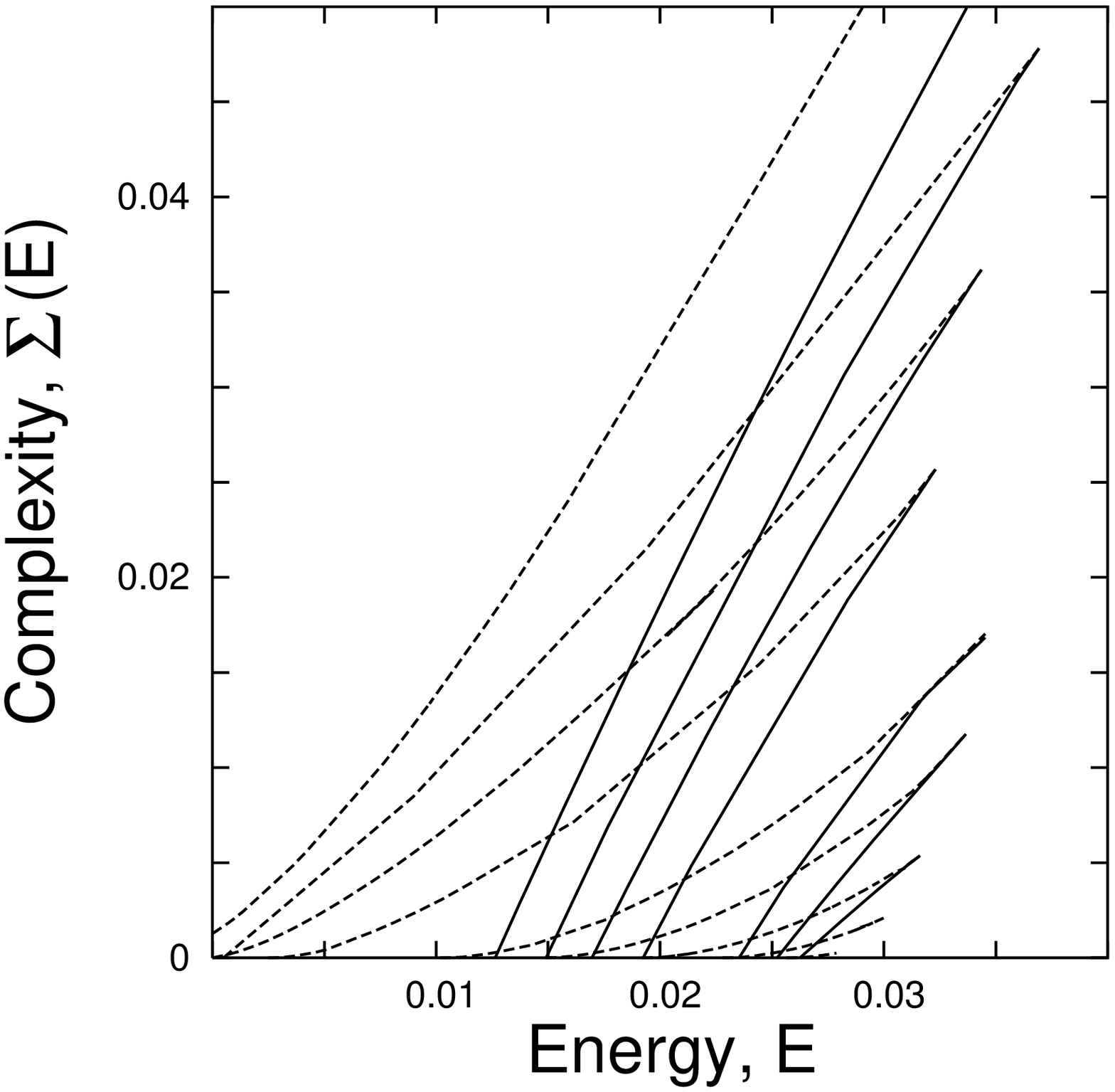}
\caption{Complexity $ \Sigma(E)$ as a function of the Ground State
 Energy $E$ for different values of the erasure probability $p$.
From right to left $p=0.606,0.613,0.62,0.63,0.64,0.66,0.67,0.68$ 
and $p=0.69$ for the code rate problem $R=1/4$.  
\label{bec_4}}
\end{figure*}

\begin{figure*}
\centering
\includegraphics*[scale=0.8]{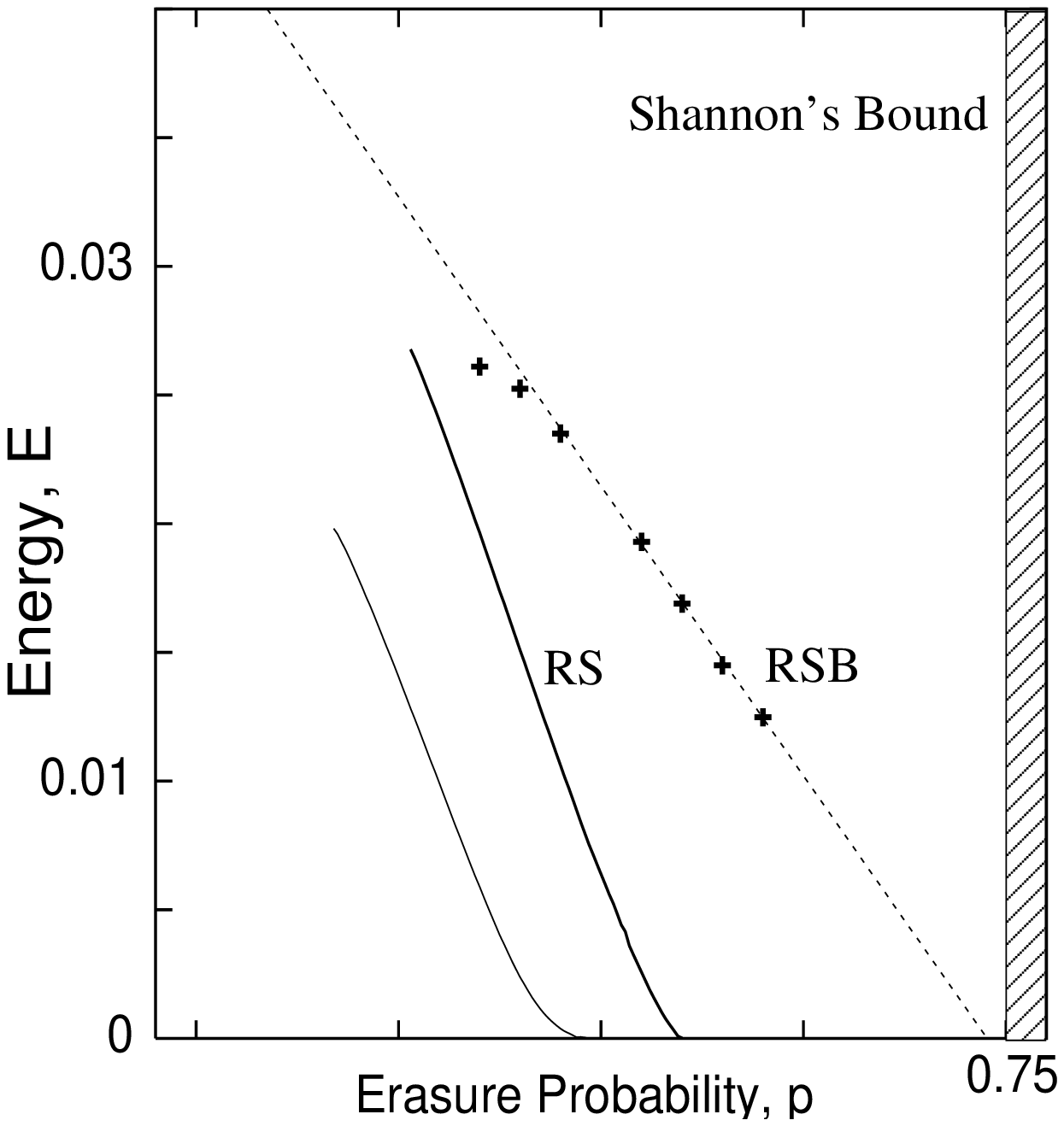}
\caption{ The RS and RSB Energy as a function of the Erasure probability.
The Shannon bound corresponding to the rate $R=1/4$ being $1-R$.
The points (bold crosses) are extrapolated from the physical branch
of the complexity at $ \Sigma=0$ and corresponds to the maximum of
the free-energy as a function of the RSB parameter
$ \mu$. The thin line corresponds to the factorized ans\"atz.
\label{bec_5}}
\end{figure*}

Moreover, considering the full convolution (\ref{crux_bec}) as we do in 
what follows, the location of the dynamical transition in the
one-step RSB solution becomes higher than that obtained within RS, 
which arguably implies that one may be able to
construct the algorithmic analog of the RSB recursion
(\ref{crux_bec}) to improve decoding performances.
Similarly, within the RSB calculation, the location of the
critical transition is increased for the rate $R=1/4$
considered above and found to be very close to the Shannon bound value $1-R =
0.75$ (cfr. Fig.~\ref{bec_5}).\\
We iterated the convolution (\ref{crux_bec}) within the one-step
RSB approximation described in section VI and computed the
free-energy as a function of the RSB 
parameter $ \mu$ for several values of the erasure probability
$p$, above the dynamical transition point $p_d \simeq 0.602$
as reported in Fig.~\ref{bec_3}. In this range of $p$ values, the
free-energy grows as a function of $ \mu$, dropping to the
ferromagnetic free-energy value (being $f_{ \mbox{\footnotesize
Ferro}}=0$ for the BEC) at some critical $ \mu^{*}$. At higher 
values of the erasure probability $p$, the free-energy
eventually drops to zero after having reached its maximum (see
Fig.~\ref{bec_3}). According to the standard recipe of maximizing the
free-energy with respect to $\mu$~\cite{MPV}, we extrapolate the
value of the dynamical transition within the one-step RSB
calculation, finding the value of $p_d^{RSB} \simeq 0.621$,
indicated in Fig.~\ref{bec_2} by the dotted line, well above the RS
value $p_d^{RS} \simeq 0.602$. The corresponding
complexity is shown in Fig.~\ref{bec_4}, for different values of the erasure
probability considered. We observe two branches, a physical (right) and an
unphysical one (left) as explained in ref.~\cite{MP2}. From the 
former branch we extrapolated the RSB energy value corresponding
to the noise channel probability. Notice that above the RS dynamical
transition we do observe, for values of $p \simeq
0.606,0.613,0.62$ that the second branch,
that is the relevant one, is not present. This can be explained because the
corresponding curves for the free-energy (Fig.~\ref{bec_3}) do not reach a
maximum and drop instead to the ferromagnetic energy value
($f=0$). The dynamical transition within the RSB calculation 
increases. 

If one considers the free-energy as a function of the
RSB parameter $ \mu$ at values of the
erasure probability slightly higher that the RS critical 
transition point, one finds that the maximum is located at positive energies,
implying that the critical transition point also changes under the
one step RSB calculation. In this way
(Fig.~\ref{bec_5}) we extrapolated the value of the critical transition, within
the RSB approximation, to $p^{RSB}_c \simeq 0.7450 \pm 0.05$. Note
that Shannon's bound for the specific code rate considered is
$1-R=3/4$. In comparison, the critical RS transition point $p^{RS}_c
\simeq 0.6695$. The fact that the
dynamical and critical transition points increase within the
one-step RSB calculation might be surprising at first. In the next
section we will show that a similar situation is observed for the
BSC at temperatures below the Nishimori condition~\cite{Nishi1},
where it is known that the RS solution is 
dominant ~\cite{NS}.
It is also reasonable to believe
that the above one step RSB calculation is exact for this problem
~\cite{MP2}~\cite{FLZ}. The above findings concerning
the location of the critical (RSB) transition seem to support
this belief. 

\subsection{The Binary Symmetric Channel}
\label{sec:BSC}
\begin{figure*}
\centering
\includegraphics*[scale=0.6]{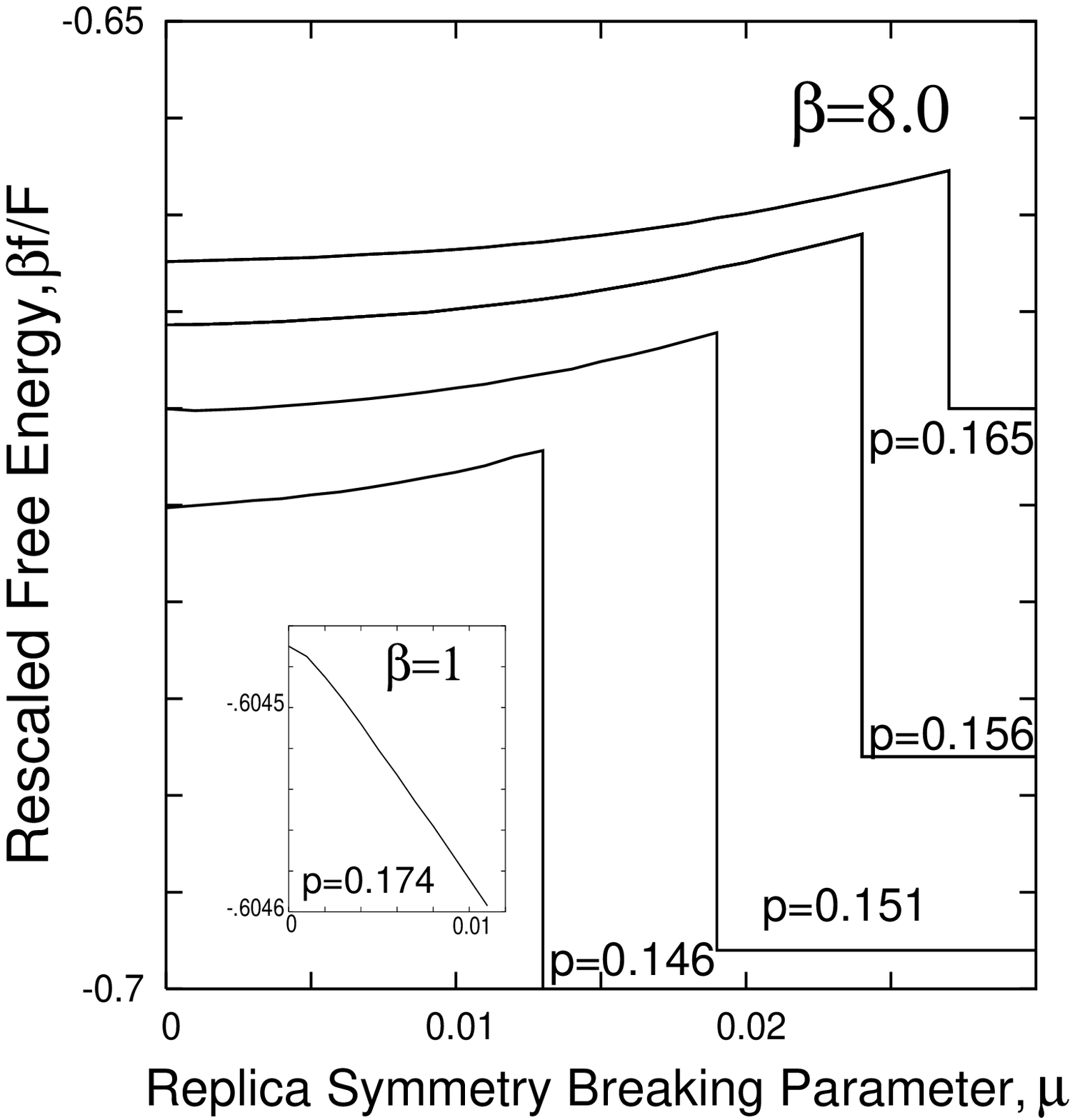}
\caption{The Gallager free-energy as a function of the Replica Symmetry
 Breaking parameter $ \mu$ for different values of the noise channel $p$
 and temperature (the inset showing the free-energy at the Nishimori
 temperature for a flip rate $p=0.174$, above the dynamical transition point.)
\label{bsc_1}}
\end{figure*}

\begin{figure*}
\centering
\includegraphics*[scale=0.7]{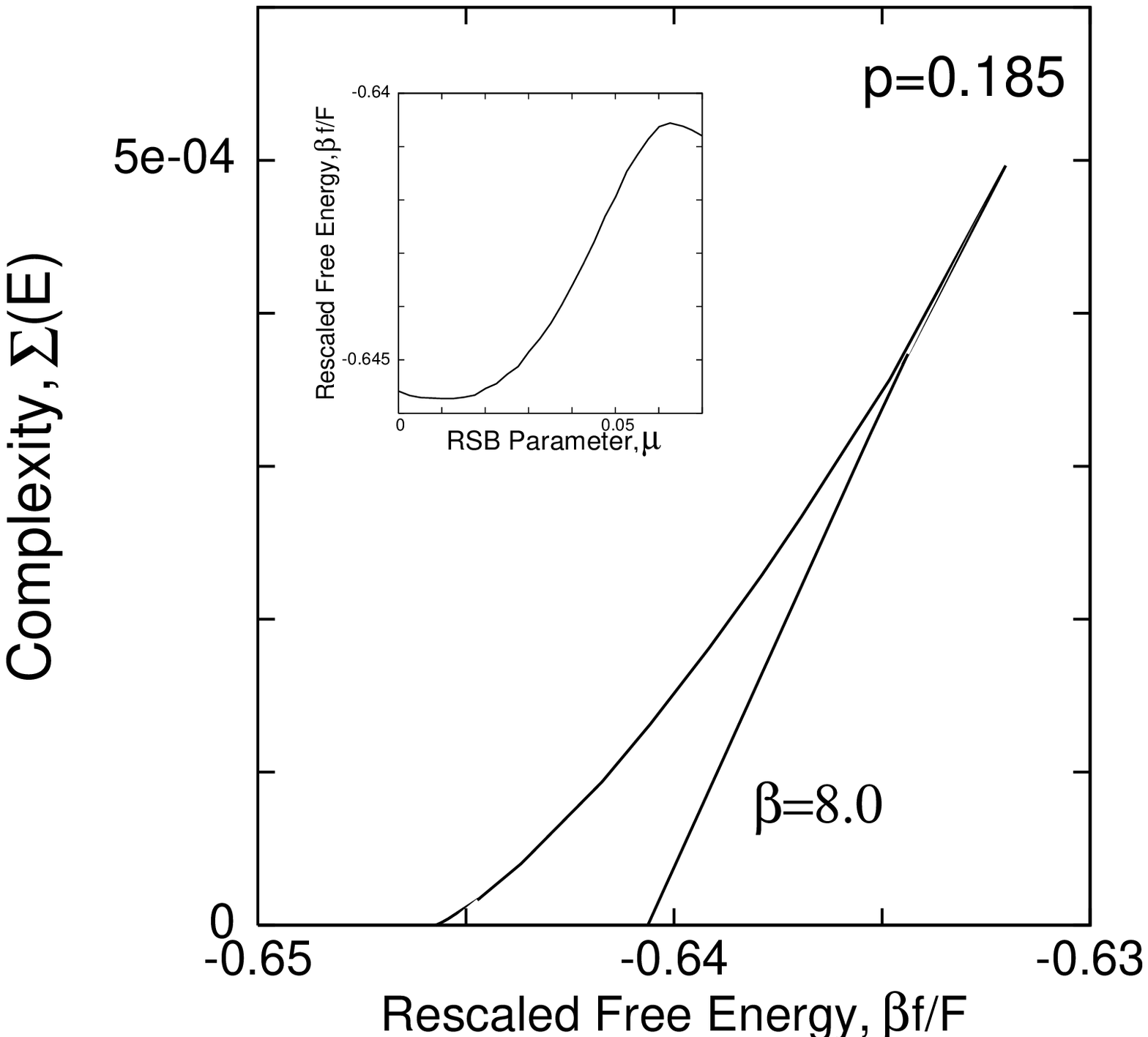}
\caption{The Gallager complexity for a BSC noise level of $p=0.185$, 
at temperature $ \beta=8.0$, below the Nishimori line. The inset shows 
the corresponding free-energy as a function of the RSB parameter $ \mu$.
\label{bsc_2}}
\end{figure*}

In this section we return to the BSC (\ref{RFRB}) and employ
the same considerations discussed for the BEC in the previous section. 
We consider the convolution
(\ref{crux_rsb}) along the Nishimori line $ \beta=1$. The location
of the dynamical transition has been discussed in
section \ref{sec:RS} and we found the value of $p^{RS}_d=0.1665$ within the
RS approximation, in agreement with probabilistic
decoding estimates. Not surprisingly~\cite{NS}, if one looks at
the free-energy as a function of the RSB parameter $\mu$ for
values of the BSC noise probability $p>p_d$, one finds that the free-energy
decreases (see the inset in Fig.~\ref{bsc_1}), indicating that the RS
solution is dominant, so that the dynamical transition is
the same in the one-step RSB case.
$p^{RS}_d=p^{1RSB}_d$. However, at temperatures well below the 
Nishimori line instead, i.e. $ \beta=8.0$, we observe the following facts:
above the RS dynamical transition $ p_d^{RS} \simeq 0.135$ the 
free-energy increases as a function of the RSB parameter $ \mu$, 
dropping to the corresponding ferromagnetic free-energy at some 
value $ \mu^{*}$, before reaching a maximum. This means that the 
second, physical branch of the complexity is not present, the 
analysis following the same lines as for the BEC calculation. 
At higher values of the flip rate, i.e. $ p \simeq 0.18$ the 
second branch of the complexity (see Fig.~\ref{bsc_2}) appears, indicating 
a non-zero complexity and that we are above the RSB dynamical 
transition. The corresponding 
complexity for the LDPC decoding at $p=0.185$ is reported in 
Fig.~\ref{bsc_2}. These findings imply that the location of the dynamical
transition  is improved with respect with the RS value.

The same behavior has been observed, progressively
reducing in importance at higher values of the temperature $T=2/8$
and $T=4/8$, while at $T=1$, corresponding to the Nishimori line,
this phenomenon is no longer observed. The inset of Fig.~\ref{bsc_1} shows
the computed free-energy as a function of the RSB 
parameter $ \mu$ at the Nishimori temperature
channel noise $ p=0.174$. 
We conclude observing that,
below the Nishimori line, the reentrant nature of the dynamical
transition is not observed within the RSB calculation.

\vspace{1cm}

\section{Conclusions and Future Perspectives.}
\label{sec:disc} 
We introduced a new method to compute the phase
diagram of finite-connectivity SG model systems and discussed the
connection with LDPC codes. We extended the
one-step RSB theory, originally formulated for the Bethe SG
problem, to the Gallager LDPC decoding for two
different type of corruption processes, BEC and BSC. For the BEC
at zero temperature we found that the location of the
dynamical transition increases with respect to the 
RS case. Our estimate for the RSB critical transition value
is a mere one percent away from capacity, differently from the RS
estimate. The critical and
dynamical properties of LDPC decoding have also been
studied for the case of a BSC. The reentrant nature of the
LDPC phase diagram and the corresponding changes that RSB
implies in its low temperature properties have also been
studied. We found that below the Nishimori line the dynamical
transition increases within the RSB theory. Possible
implications to improve the decoding performances of the
state-of-the-art error correcting codes are one of the
perspectives we wish to address. We would like to
construct the algorithmic analog of the convolution
(\ref{crux_rsb}) in order to improve the decoding performances as
predicted by the RSB theory. The method we used to determine the
phase diagrams along this work is very promising. It allows to
study, without loss of continuity, rather different problems,
ranging from the $d$-dimensional hierarchical SG to finite-connectivity
mean-field models. In this sense it would be very interesting to
apply this method to clarify a few open issues. On the one side we
would like to investigate, within the one-step RSB theory, the
finite temperature phase diagram of the Bethe SG in a magnetic field. On the
other, it would be interesting to explore the presence, if any, of
RSB phenomena in $d$-dimensional SG hierarchical model systems.

\begin{acknowledgments}
This work was supported in part by the European Community's Human
Potential Programme under contract number HPRN-CT-2002-00319.
\end{acknowledgments}

\section*{Appendix A - Equivalence of the RS Equations}

Considering the replicated partition function,
\begin{eqnarray}
\langle {\cal Z}^n \rangle_{ {\cal A}, \zeta,J} = E_{ { \cal A},
J, \zeta}
Tr_{ \{S_j^{ \alpha}\}} \times \nonumber \\
e^{ \beta F \sum_{\alpha,k} \zeta_k S_{k}^{\alpha}+ \beta \sum_{
\alpha,k} { \cal A}_{\mu} J_{ \mu}
S_{i_1}^{ \alpha} \cdots S_{i_K}^{ \alpha} } \nonumber \\
\label{partition_f}
\end{eqnarray}
one can perform the average over the corresponding underlying
geometry in (\ref{partition_f}) introducing an integral
representation for the finite-connectivity constrain $ \delta (
\sum_{ \mu \setminus i} { \cal A}_{ \mu} -C)$ in the spirit
of~\cite{WS}, defining the set of overlaps $q_{ \alpha_1 \cdots
\alpha_m }$ and conjugated functional order parameter $ \hat{q}_{ \alpha_1
\cdots \alpha_m}$, ~\cite{Saad}. This leads, under the RS 
ans\"atz,
\begin{eqnarray}
q_{ \alpha_1 \cdots \alpha_m} &=& \int dx~ \pi(x) x^m \nonumber \\
\hat{q}_{ \alpha_1 \cdots \alpha_m} &=& \int d \hat{x} ~ \hat{ \pi}( \hat{x}) \hat{x}^m, \nonumber \\
\end{eqnarray}
to the following expression for the free energy,
\begin{eqnarray}
\beta f = &-&  \frac{C}{K} \ln \cosh \beta J
 - \frac{C}{K} E_J \int \Big [ \prod_{l=1}^K dx_l \pi(x_l) \Big ]
  \nonumber \\
& \times & \ln \Big [ 1+ \tanh( \beta J ) \prod_{j=1}^K
\tanh( \beta x_j) \Big ] \nonumber \\
&+& C \int dx d \hat{x} \pi(x) \hat{ \pi} (\hat{x})
\log \big ( 1+\tanh( \beta x) \tanh( \beta \hat{x}) \big ) \nonumber \\
&+& C \int dy \hat{ \pi} (\hat{x}) \log \cosh ( \beta \hat{x}) \nonumber \\
&-& E_{ \zeta } \int \prod_{l=1}^C dy_l \hat{ \pi}(x_l)  \Big [ 2
\cosh
 \beta ( \sum_{l=1}^C \hat{x}_l +  F \zeta ) \Big ].
\label{free-energy2}
\end{eqnarray}
The corresponding saddle point equations (\ref{crux2}) are derived
differentiating the above expression with respect to the local field
$ \pi(x)$ and its conjugated distribution $ \hat{ \pi} ( \hat{x}
)$. In order to obtain the free-energy (\ref{free-energy}), the
second and third term in (\ref{free-energy2}) can be combined via
equation (\ref{crux2}), to obtain
\begin{eqnarray}
II+III \equiv &+& C \frac{K-1}{K} \Delta f^{(2)} - C \frac{K-1}{K} \log \cosh \beta J
 \nonumber \\
 &-& C (K-1) \int dx ~\pi(x) \log \cosh ( \beta x).\nonumber \\
\end{eqnarray}
The fourth term can be written as
\begin{eqnarray}
IV \equiv \Delta f^{(1)}_b &+& C \log \cosh ( \beta J) \nonumber \\
&+& C(K-1) \int dx ~\pi(x) \log \cosh ( \beta x), \nonumber \\
\end{eqnarray}
where we wrote $\Delta f^{(1)} = \Delta f^{(1)}_a+\Delta
f^{(1)}_b$ and where the last term in the free-energy expression
above is nothing but $\Delta f^{(1)}_a$. It is easy to see that
several terms cancels out, together with the first term in
equation (\ref{free-energy2}), to obtain equation
(\ref{free-energy}). Similar simplifications and their
interpretation were discussed already in the case of the Bethe SG
\cite{MP2}.

\section*{Appendix B -The RSB equations }

Consider the set of coupled equations ( \ref{crux2}),
\begin{eqnarray}
\pi(x)&=& E_{ \zeta} \int \prod_{i=1}^{C-1} d \hat{x}_i~ \hat{ \pi
}(\hat{x}
 )~ \delta \Big ( x - \sum_{i=1}^{C-1} \hat{x_i}- F \zeta \Big) \nonumber \\
\hat{ \pi }( \hat{x} ) &=& E_J \int \prod_{j=1}^{K-1} dx_j ~\pi(x_j) ~ \delta
\Big( \hat{x}-u( \{ x_j\},J) \Big).
\label{crux2b}
\end{eqnarray}
For the case of $K=2$, with no random bias,
corresponding to the Bethe SG problem, let us define the Laplace
transform of the conjugated local field distribution as
\begin{equation}
g( \sigma )= \int d \hat{x}~ e^{ - \sigma \hat{x} } \hat{ \pi}(\hat{x} ) .
\label{gop2}
\end{equation}
Substituting the first equation of (\ref{crux2b}) into the second and
then into ( \ref{gop2}), one finds
\begin{eqnarray}
g ( \sigma ) = \int ds ~ \delta( s+ i \sigma)~ E_{ J, \zeta}
\int \prod_{j=1}^{C-1} d \hat{x}_j~ \hat{ \pi }( \hat{x}_j ) \times \nonumber \\
\exp \left[ \frac{i}{ \beta}  \tanh^{-1} ( \tanh( \beta J) \tanh ( \beta
      \sum_{j=1}^{C-1} \hat{x}_j + \beta F ) ) \right ]. \nonumber \\
\end{eqnarray}
which is the expression for the global order parameter
found in~\cite{Mo} and corresponds to expression (\ref{RS}) for
$K=2$. The expression for arbitrary $K$ values of the multi-spin
interaction simply involves multiple convolutions of the Mottishaw
global order parameter. The meaning of the global order parameter
(\ref{gop2}) can be understood observing that, e.g. in the case
of the Bethe SG considered here (in a uniform magnetic field), the
partition function is given by~\cite{Mo}
\begin{equation}
\langle Z^n \rangle \equiv Tr_{ \{ \sigma_o^{ \alpha} \} } \exp \Big [ \beta F \sum_{ \alpha=1}^n
\sigma_o^{ \alpha} \Big ] [g_m( \sigma_o^{ \alpha }) ]^C,
\end{equation}
where
\begin{eqnarray}
g( \{ \sigma_o^{ \alpha } \} )= E_J \Big [ Tr_{ \{ \sigma_1^{ \alpha } \} }
(g( \{ \sigma_1^{ \alpha } \} ))^{C-1} \nonumber \\
\exp \Big [ \beta J \sum_{ \alpha} \sigma_1^{ \alpha} \sigma_o^{ \alpha}
+ \beta F \sum_{ \alpha} \sigma_1^{ \alpha} \Big ] \Big ]. \nonumber \\
\label{RSB2}
\end{eqnarray}
The subscript $m$ indicates the dependence of the global order
parameter on the sub-shell tree structure. If one is interested in
the thermodynamical limit, it is expected that the fixed point of
the above recursion determines the equilibrium properties of the
system. A delicate point is the choice of the boundary conditions
which are well known~\cite{MP2} to play an important role in this
problem. The corresponding extension of equation (\ref{RSB2}) to
the case of arbitrary $K$ values has been reported in equation
(\ref{gopK}).

\end{document}